\documentclass[twocolumn,showpacs,preprintnumbers,amssymb,aps,pra,a4paper]{revtex4-1}
\usepackage{graphicx,rotating}
\usepackage{epsfig}
\usepackage{dcolumn}
\usepackage{bm}
\usepackage{amssymb}
\usepackage{latexsym,epsfig}
\usepackage{amsmath}
\usepackage{color}
\usepackage{amstext}
\usepackage{tikz}

\newcommand{\circlesign}[1]{
    \mathbin{
        \mathchoice
        {\buildcirclesign{\displaystyle}{#1}}
        {\buildcirclesign{\textstyle}{#1}}
        {\buildcirclesign{\scriptstyle}{#1}}
        {\buildcirclesign{\scriptscriptstyle}{#1}}
    }
}

\newcommand\buildcirclesign[2]{%
    \begin{tikzpicture}[baseline=(X.base), inner sep=0, outer sep=0]
    \node[draw,circle] (X)  {\ensuremath{#1 #2}};
    \end{tikzpicture}%
}

\begin{document}

\title{Diverse Trends of Electron Correlation Effects for Properties with Different Radial and Angular Factors in an Atomic System: A case study in Ca$^{+}$}
\vspace*{0.5cm}

\author{$^{a,b}$Pradeep Kumar, $^c$Cheng-Bin Li and $^{a,c}$B. K. Sahoo}
\affiliation{$^a$Atomic, Molecular and Optical Physics Division, Physical Research Laboratory, Navrangpura, Ahmedabad-380009, India\\
$^b$Indian Institute of Technology Gandhinagar, Ahmedabad, India\\
$^c$State Key Laboratory of Magnetic Resonance and Atomic and Molecular Physics, Wuhan Institute of Physics and Mathematics,
Chinese Academy of Sciences, Wuhan 430071, China}

\email{Emails: pradeep@prl.res.in; cbli@wipm.ac.cn; bijaya@prl.res.in}

\date{Received date; Accepted date}
\vskip1.0cm

\begin{abstract}
Atomic properties such as field shift constants, magnetic dipole and electric quadrupole hyperfine structure constants, Land\'{e} $g_J$ factors,
and electric quadrupole moments that are described by electronic operators with different ranks and radial behaviors are studied and the role
of electron correlation effects in their determination are investigated. We have adopted the Dirac-Hartree-Fock method, the second- and third-order 
relativistic many-body perturbation theories, and an all-order relativistic many-body method in the coupled-cluster theory framework considering only 
the linearized terms and also all the non-linearized terms in the singles and doubles with partial triples excitations approximation to carry out
these analyses. Variations in the propagation of electron correlation effects with operators having same angular factors but different radial
behaviors and with different ranks are highlighted. Corrections from the higher-order relativistic corrections due to the Breit and quantum electrodynamics
interactions to all these properties are also estimated. Understanding of trends of electron correlation effects in these properties can be useful
to establish accuracies in the theoretical results of different atomic properties and to substantiate validity of an approximated many-body method.
\end{abstract}
\pacs{31.15.A-, 31.15.B-, 31.15.V-, 31.30.Gs}
\maketitle

\section{Introduction}

Investigating role of electron correlation effects in many-electron systems such as atoms, molecules and solids are the long standing
research problems~\cite{bartlett}. Electronic wave functions of these systems with more than three electrons are not practical to solve exactly, hence
approximated methods are being used for their calculations. The first-step in these approaches is to treat a mean-field or effective
potential to obtain approximated wave functions and energies. In the atomic systems, Hartree-Fock (HF) method in the non-relativistic
case or Dirac-Hartree-Fock (DHF) method in the relativistic framework is being considered as the most suitable mean-field approach to obtain
the approximated wave functions and energies as it is devised on the basis of variational principle~\cite{szabo,helgaker}. Going beyond the (D)HF
method for more accurate calculations, one uses finite-order many-body perturbation theory (MBPT) or all-order perturbative methods truncating at
some levels of approximations depending upon the choice or existing computational facilities.

 There have been continuous attempts to benchmark the truncated many-body methods for producing accurate theoretical results in various perspectives.
For example, an accurate many-body method can predict reliable spectroscopic data on the systems which are not experimentally observed. Very
accurate matrix elements of weak-interaction Hamiltonians are required to find out exotic physics from the studies of atomic parity
non-conservation (PNC) effects \cite{aoki,bksahoo1,ginges} and permanent electric dipole moments (EDMs) of the atoms \cite{ginges,yamanaka}. Thus,
it would be imperative to establish capability of a many-body method so that it can be employed assuredly for theoretical studies in guiding many 
high-precision experiments such as atomic clocks \cite{bksahoo2}, quantum information studies \cite{nielsen}, probing violations of discrete
symmetries \cite{ginges,yamanaka}, Lorentz symmetry violation \cite{dzuba}, etc. prior to performing the actual measurements.

There have also been immense interest to acquire insightful knowledge on the behavior of electron correlation effects in atomic systems. In the
Slater determinant formalism, it is obvious to comprehend that a full configuration interaction (CI) method can account for all the correlation
effects in a system but this method is not viable to apply to the atomic systems having more than four electrons. Again, though this method can
give final results that can be treated as the numerical experimental values, it cannot demonstrate the roles of electron correlation effects explicitly.
Alternative all-order methods such as Green's function approach and relativistic coupled-cluster (RCC) methods, that are formulated adapting
diagrammatic techniques, are apt to demonstrate roles of different correction effects categorically \cite{szabo,helgaker,bartlett}. Moreover, when
these methods are truncated at different levels of approximation and employed to atomic systems, one could see contrasting results owing to
limitations over capturing electron correlation effects through various methods. Depending upon the {\it ansatz} followed to express the form of
the wave functions, the truncated many-body methods can have many advantages or disadvantages in capturing various correlation effects and can
claim for extensive computational resources. On this point of view the truncated RCC methods, let's say at the singles and doubles excitations
approximation (CCSD method), is more powerful than the truncated CI method at the same truncation level of excitations \cite{szabo,bartlett}. As a
matter of fact, the singles and doubles CI approximation (CISD method) has size consistency and size extensivity problems, whereas the CCSD
method is free from them \cite{szabo,bartlett}. In fact, the CCSD method
can account for contributions from the higher level excitations such as triples, quadruples etc. to some extent through its non-linear terms. However,
one can find an approximated multi-configuration Dirac-Fock (MCDF) method, which is a variant of CI method, can produce sometimes more accurate
spectroscopic data than the CCSD method \cite{tommaseo,itano}. On the other hand, the CCSD method is proven to give some of the results more
precisely, while other calculations are far off from the experimental results \cite{bijaya1,bijaya2}.

As has been mentioned earlier, there are colossal demands to authorize methods that can offer very reliable results in atomic systems so that
they can be used to determine quantities that cannot be yield from direct measurements. In studies on PNC and EDM effects in atomic systems,
it is imperative to ascertain accuracies of the theoretical results to accomplish their objectives \cite{aoki,bksahoo1,ginges,yamanaka}. These
calculations require meticulously determinations of the electric dipole (E1) matrix elements, excitation energies and matrix elements of the PNC
and EDM interaction Hamiltonians between different atomic states \cite{bksahoo1,yamanaka}. In principle, accuracies of the E1 matrix elements can
be verified by estimating lifetimes or E1 polarizabilities ($\alpha_d$) of atomic states and comparing them with the corresponding measurements.
Analogously, accuracies of the excitation energies can be justified by comparing the calculated values from a method with their experimental values.
However, such approaches cannot be adopted to gauge accuracies of the atomic PNC and EDM interaction Hamiltonians. Alternative approaches like
comparing calculated magnetic dipole (M1) hyperfine structure constants ($A_{hf}$) and $\alpha_d$ values with the experimental results are being
commonly used to find out accurate behaviors of the wave functions in the nuclear region \cite{bks1} and to test the validity of the methodology
employed in the calculations, respectively \cite{bks2,bks3}. It is, however, dubious to justify whether it is really able to estimate accuracies
of the matrix elements of a particular operator by analyzing matrix elements of another operator having same rank but different radial behavior
or vice versa. Thus, this needs to be ascertained by carrying out rigorous analysis before acclaiming such justifications.

There are also similar perspectives are undertaken to gauge accuracies of some of the properties by determining other properties. To name a few,
accurate evaluations of isotope shifts, particularly field shifts, are justified by calculating the $A_{hf}$ values as both the quantities are sensitive
to the behaviors of the atomic wave functions in the nuclear region \cite{martensson,wendt,brage}. It should be noted that ranks of the operators
associated with these evaluations are not same and radial dependencies of these properties are also different. On the other hand, the single particle
matrix elements of the electronic components of the $A_{hf}$ and Land\'{e} $g_J$ factor determining operators have the same angular factors but their
radial dependencies are very contrast \cite{bksahoo2,cheng}. Also, electronic components of the electric quadrupole (E2) hyperfine structure
constant ($B_{hf}$) determining operator and the electric quadrupole moment operator share the same angular factors while differ in their radial
dependencies. Thus, it would be pragmatic to fathom propagation of electron correlation effects in all these properties in a particular atomic
system employing many-body methods that are approximated at different levels to acquire comprehensive understanding about their capabilities on
producing theoretical results matching with the available experimental results. This may enlighten us to learn more about credentials of the
approximated methods that are usually employed for performing spectroscopic properties in an atomic system. With this objective, we carry out
investigation of trends of electron correlation effects in evaluating a variety of atomic properties that are described by physical operators
having different ranks and radial dependencies. We also employ relativistic many-body methods with various approximations in order to substantiate 
competence of these methods for attaining reliable results. For this analysis, we choose the singly charged calcium ion (Ca$^+$), as it has been 
undertaken for many high precision experiments such as for atomic clock \cite{huang,matsubara}, quantum information \cite{chwalla}, probing Lorentz 
symmetry violation \cite{pruttivarasin}, investigating field-shift ratio \cite{shi}, measuring $g_J$ factor \cite{tommaseo,chwalla}, etc. by many 
groups around the globe.

\section{THEORY}\label{sec2}

For the intended investigation, we would like to consider even parity operators with different radial and angular momentum dependencies
for calculating their expectation values employing a number of truncated many-body methods to demonstrate propagation of the electron correlation
effects systematically. So we choose specifically operators describing $A_{hf}$, $B_{hf}$, $g_J$ factor, field-shift constant ($F$) and
electric quadrupole moment ($\Theta$) for this objective. Clearly, they are the scalar, vector and tensor type operators ranging ranks from 0 to 2.
Moreover, some of these operators have same angular momentum but different radial dependencies. Therefore, a comparative analysis in the trends of the
electron correlation effects can be articulated by investigating these properties in a system using various approximated many-body methods. To
make it evident, we give the single particle matrix element expressions of the respective operators below.

\subsection{Field shift constant $F$}

Conventionally, field-shift is theoretically determined assuming uniform charge distribution inside the atomic nucleus in a sphere of radius $R$. This 
can yield expression for the field-shift determining operator $F_{fs}$ as \cite{johnson}
\begin{eqnarray}
F_{fs}=\sum_i f_{fs}^{(0)} (r_i) &=& \begin{cases} \frac{5Z}{4R^3}\left[1-\frac{r_i^2}{R^2}\right] & \text{for} \ \ r \le R  \\
          0 & \text{for} \ \ r > R,
         \end{cases}
\label{eqn1}
\end{eqnarray}
where $Z$ is the atomic number. The reduced matrix element of $f_{fs}^{(0)}$ is given by
\begin{eqnarray}
\langle \kappa_f||f_{fs}^{(0)}||\kappa_i\rangle &=& \langle \kappa_f||\circlesign{C^{(0)}}||\kappa_i\rangle \int^{\infty}_{0}dr \circlesign{f_{fs}^{(0)}(r)}
 \nonumber \\ && \times \left (P_{f}P_{i}+Q_{f}Q_{i} \right ),
 \label{eqn2}
\end{eqnarray}
where $P$ and $Q$ are the radial parts of the large and small components of the single particle Dirac wave function, respectively, $\kappa$ is the
relativistic angular momentum quantum number and $C^{(k)}$ is the Racah operator of rank $k$. In the above expression, we have encircled quantities
that define rank of the operator and that is responsible for radial dependency apart from the wave function components. Since $F_{fs}$ is finite only within
the nucleus, investigating $F_{fs}$ can test the accuracies of the wave functions in the nuclear region. The reduced matrix element of the
Racah tensor $C^{(k)}$ is given by
\begin{eqnarray} \nonumber
 \langle k_{f}||C^{(k)}||k_{i}\rangle&=&(-1)^{j_{f}+1/2}\sqrt{(2j_{f}+1)(2j_{i}+1)}\\
 &&\times\begin{pmatrix}
  j_{f} & k & j_{i}\\
  1/2 & 0 & -1/2 \\
\end{pmatrix}\pi(l_{f},k,l_{i}),
\label{eqn3}
\end{eqnarray}
with the selection rule
\begin{eqnarray}
\pi(l_{f},k,l_{i})=\begin{cases}
1,  \hspace{0.3cm}\text{if}\hspace{0.3cm} l_{f}+k+l_{i}\hspace{0.3cm} \text{even}\\
0,  \hspace{0.3cm}\text{otherwise}\\
\end{cases}
\label{eqn4}
\end{eqnarray}
where $l$ is the orbital angular momentum of the corresponding orbital. 

The field shift constant is defined as $F= \langle F_{fs} \rangle$ for a given atomic state. 

\subsection{Land\'{e} $g_J$ factor}

The Dirac contribution to the $g_{J}^D$ factor of a bound electron can be evaluated by \cite{cheng}
\begin{eqnarray}
g_{J}^D=\frac{\langle J||\textbf{N}^{(1)}||J \rangle}{2\mu_{B}\sqrt{J(J+1)(2J+1)}},
 \label{eqn5}
\end{eqnarray}
where $J$ is the total angular momentum of the atomic state being considered and $\textbf{N}^{(1)} = \sum_i \mu^{(1)}_{q}(r_i)$ such as the
corresponding single particle reduced matrix element
of $\mu^{(1)}$ is given by
\begin{eqnarray}
\langle \kappa_{f}||\mu^{(1)}||\kappa_{i}\rangle &=&-(\kappa_{f}+\kappa_{i})\langle -\kappa_{f}|| \circlesign{C^{(1)}} ||\kappa_{i}\rangle \nonumber \\
&&\times\int^{\infty}_{0}dr \circlesign{ \ r \ } \left (P_{f}Q_{i}+Q_{f}P_{i} \right) .
\label{eqn6}
\end{eqnarray}
As seen from the above equation, this quantity is directly proportional to the radial distance of the electron.

 The quantum electrodynamic (QED) correction to the $g_J^D$ factor ($\Delta g_J^Q$) can be estimated approximately using the
expression \cite{cheng}
\begin{eqnarray}
\Delta g_{J}^Q=0.001160 \frac{\langle J||\Delta\textbf{N}^{(1)}||J \rangle}{\sqrt{J(J+1)(2J+1)}},
 \label{eqn7}
\end{eqnarray}
with $\Delta \textbf{N}^{(1)} = \sum_i \Delta \mu^{(1)}_{q}(r_i)$ for the reduced matrix element
\begin{eqnarray}
\langle \kappa_{f}||\Delta\mu^{(1)}|| \kappa_{i}\rangle=-\langle \kappa_{f}||C^{(1)}||\kappa_{i}\rangle\int^{\infty}_{0}dr
(P_{f}P_{i}+Q_{f}Q_{i}).
\label{eqn8}
\end{eqnarray}
Thus, the net value is given by $g_J=g_J^D+\Delta g_J^Q$. The electron correlation effects hardly play any role in estimating the $\Delta g_J^Q$
correction \cite{pradeep}, so the angular and radial dependencies of the expression given by Eq. (\ref{eqn5}) is reflected in the determination of
the $g_J$ factor through different many-body methods.

\subsection{Hyperfine structure constant $A_{hf}$}

The M1 hyperfine structure constant is expressed as \cite{schwartz}
\begin{eqnarray}
 A_{hf}=\mu_{N} g_{I} \frac{\langle J||{\bf T}_{hf}^{(M1)}||J\rangle}{\sqrt{J(J+1)(2J+1)}},
 \label{eqn9}
\end{eqnarray}
where $\mu_{N}$ is the nuclear magneton, $g_{I}$ is the ratio of nuclear magnetic dipole moment $\mu_{I}$ and the nuclear spin $I$.
The single particle matrix element of the M1 hyperfine interaction operator $T_{hf}^{(M1)}=\sum_i t_{hf}^{(1)}(r_i)$ is given by
\begin{eqnarray}
\langle \kappa_{f}||t_{hf}^{(1)}||\kappa_{i}\rangle &=&-(\kappa_{f}+\kappa_{i})\langle -\kappa_{f}||\circlesign{{\bf C}^{(1)}}||\kappa_{i}\rangle \nonumber \\
&& \times \int^{\infty}_{0}dr \frac{(P_{f}Q_{i}+Q_{f}P_{i})}{\circlesign{r^{2}}} .
\label{eqn10}
\end{eqnarray}
 Obviously, this quantity is sensitive in the nuclear region owing to the inversely proportional to square of the radial distance of the
 electrons. In the present analysis, we use the value as $g_{I}=-0.37646943$ of $^{43}$Ca$^+$ \cite{stone}.

\subsection{Hyperfine structure constant $B_{hf}$}

The E2 hyperfine structure constant is expressed as \cite{schwartz}
\begin{eqnarray}
B_{hf}&=&Q_{nuc}\left\lbrace\frac{8J(2J-1)}{(2J+1)(2j+2)(2j+3)}\right\rbrace^{1/2} \nonumber \\&& \times \langle J||{\bf T}_{hf}^{(2)}||J\rangle,
\label{eqn11}
\end{eqnarray}
where $Q_{nuc}$ is the nuclear quadrupole moment. The single particle matrix element of the E2 hyperfine
interaction operator $T_{hf}^{(E2)}=\sum_i t_{hf}^{(2)}(r_i)$ is given by
\begin{eqnarray}
\langle \kappa_f ||t_{hf}^{(2)}||\kappa_i \rangle &=& -\langle \kappa_f||\circlesign{C^{(2)}}||\kappa_i \rangle \nonumber \\
&& \times \int^{\infty}_{0} dr \frac{(P_{f}P_{i}+Q_{f}Q_{i})}{\circlesign{r^3}}.
\label{eqn12}
\end{eqnarray}
This quantity is more sensitive in the nuclear region for its $1/r^3$ dependency. We have considered $Q_{nuc}=0.0444(6)b $ of $^{43}$Ca$^+$
\cite{sahoo1} for the theoretical determination of $B_{hf}$ values.

\subsection{Electric quadrupole moment $\Theta$}

The $\Theta$ value of an atomic state is the expectation value of the quadrupole operator given by \cite{itano}
\begin{eqnarray}
\mathcal{Q}_{el} = \sum_{i} q^{(2)}_{m} (r_i) = -\frac{e}{2}\sum_{i}(3z^{2}_{i}-r^{2}_{i}),
\label{eqn13}
\end{eqnarray}
with the single particle reduced matrix element given by
\begin{eqnarray}
\langle \kappa_f ||q^{(2)}||\kappa_i \rangle &=& \langle \kappa_f || \circlesign{C^{(2)}} ||\kappa_i \rangle \int^{\infty}_0
dr \circlesign{r^2} \nonumber \\ && \times (P_{f}P_{i}+Q_{f}Q_{i}).
\label{eqn14}
\end{eqnarray}

As seen now in Eqs. (\ref{eqn2}), (\ref{eqn6}), (\ref{eqn10}), (\ref{eqn12}) and (\ref{eqn14}) all the expressions that we are going to
consider in our calculations have different angular and radial dependencies. Below we discuss some of the relativistic many-body methods
that we would like to employ for calculating the aforementioned properties in the low-lying $4s ~ ^2S_{1/2}$, $3d ~ ^2D_{3/2}$, $3d ~ ^2D_{5/2}$,
$4p ~ ^2P_{1/2}$, and $4p ~ ^2P_{3/2}$ states of $^{43}$Ca$^+$. Comparison of the foregoing properties among these states belonging to
different orbital angular momenta and parities can demonstrate variation in the correlation trends in these states as well as, it can illustrate
their dependencies with the ranks and radial behaviors of the associated operators.

\section{Methods for calculations}

The objective of this work is to investigate roles of the electron correlation effects in the evaluation of properties described by operators with
varying rank and radial dependencies in low-lying states of Ca$^+$. For this, we consider the Dirac-Coulomb (DC) interaction Hamiltonian to
calculate the atomic wave functions. The DC Hamiltonian in atomic unit (a.u.) is given by
\begin{eqnarray}
 H=\sum_{i}[c\pmb{\alpha}\cdot \mathbf{p}_{i}+(\beta-1)c^2+V_{\text{nuc}}(r_{i})]+\sum_{i\le j}\frac{1}{r_{ij}} ,
\label{eqn:dcHamiltonian}
\end{eqnarray}
where $c$ is the velocity of light, $\pmb{\alpha}$ and $\beta$ are the Dirac matrices, $V_{\text{nuc}}(r)$ is the nuclear potential evaluated
considering the Fermi charge distribution and $\frac{1}{r_{ij}}=\frac{1}{|{\bf r}_{ij}|}=\frac{1}{|{\bf r}_i - {\bf r}_j|}$ is the two-body interaction potential between
the electrons located at ${\bf r}_i$ and ${\bf r}_j$.

It is found in the previous calculation of the $g_J$ factor of the ground state of Ca$^+$ that, the frequency independent Breit interaction is
quite significant. We also estimate contributions due to this interaction by adding the corresponding interaction potential energy in the atomic
Hamiltonian as given by
\begin{eqnarray}
V_B(r_{ij})=-\frac{\{\mbox{\boldmath$ \alpha$}_i\cdot \mbox{\boldmath$ \alpha$}_j+
(\mbox{\boldmath$ \alpha$}_i\cdot {\bf \hat{r}}_{ij})(\mbox{\boldmath$\alpha$}_j\cdot {\bf \hat{r}}_{ij}) \}}{2r_{ij}} ,
\end{eqnarray}
where ${\bf \hat{r}}_{ij}$ is the unit vector along ${\bf r}_{ij}$.

We also consider the lowest-order corrections due to the vacuum potential (VP) and self-energy (SE) effects in the calculations of the wave
functions of the bound electrons as described in our previous work \cite{bkscs}, considering the nuclear Fermi charge distribution. The VP
potential is accounted for as sum of the Uehling ($V_{U}(r)$) and Wichmann-Kroll ($V_{WK}(r)$) potentials given by \cite{flambaum}
\begin{eqnarray}
V_{U}(r)&=&  - \frac{2 \alpha_e^2 }{3 r} \int_0^{\infty} dx \ x \ \rho_{\text{nuc}}(x)
\int_1^{\infty}dt \sqrt{t^2-1} \nonumber \\ && \times
\left(\frac{1}{t^3}+\frac{1}{2t^5}\right)  \left [ e^{-2ct|r-x|} - e^{-2ct(r+x)} \right ] \ \ \
\end{eqnarray}
and
\begin{eqnarray}
V_{WK}(r)&=&-\frac{8 Z^2 \alpha_e^4 }{9 r} (0.092) \int_0^{\infty} dx \ x \ \rho_{\text{nuc}}(x)  \nonumber \\ && \times \big ( 0.22
\big \{ \arctan[1.15(-0.87+2c|r-x|)] \nonumber \\ && - \arctan[1.15(-0.87+2c(r+x))] \big \} \nonumber \\ && + 0.22
\big \{ \arctan[1.15(0.87+2c|r-x|)] \nonumber \\ && - \arctan[1.15(0.87+2c(r+x))] \big \} \nonumber \\
&& - 0.11 \big \{ \ln[0.38 -0.87c|r-x|+c^2 (r-x)^2 ] \nonumber \\
&& - \ln[0.38 -0.87c (r+x) + c^2 (r+x)^2 ] \big \} \nonumber \\
&& + 0.11 \big \{ \ln[0.38 +0.87 c |r-x| + c^2 (r-x)^2 ] \nonumber \\
&& -  \ln[0.38 +0.87c (r+x) + c^2 (r+x)^2 ] \big \} \big ), \ \ \ \ \
\end{eqnarray}
respectively, where $\rho_{\text{nuc}}$ is the nuclear density. Similarly, the SE potential energy is evaluated as the sum of contributions from the
electric and magnetic form-factors ($V_{SE}(r)=V_{SE}^{ef}(r)+V_{SE}^{mg}(r)$) following the approach of Ref. \cite{flambaum}. These expressions
are given by
\begin{eqnarray}
V_{SE}^{ef}(r)&=& - A(Z) (Z \alpha_e )^4 e^{-Zr} + \frac{B(Z,r) \alpha_e^2 }{ r} \nonumber \\ 
&& \times \int_0^{\infty} dx  x  \rho_{\text{nuc}}(x) \int^{\infty}_1 dt \frac{1}{\sqrt{t^2-1}}  
\big \{ \left( \frac{1}{t}-\frac{1}{2t^3} \right ) \nonumber \\ 
&& \times \left [ \ln(t^2-1) +4 \ln \left ( \frac{1}{Z \alpha } +\frac{1}{2} \right ) \right ] \nonumber \\
&& -\frac{3}{2}+\frac{1}{t^2} \big \} \left [ e^{-2ct|r-x|} - e^{-2ct(r+x)} \right ]
\end{eqnarray}
and from the magnetic form-factor given by
\begin{eqnarray}
V_{SE}^{mg} (r) &=& \frac{i \alpha_e }{4 \pi c} \mbox{\boldmath$\gamma$} \cdot \mbox{\boldmath$\nabla$}_r \int_0^{\infty} d^3 x \ \rho_{\text{nuc}}(x)
\nonumber \\ && \times \left [ \left ( \int^{\infty}_{1}dt \frac{e^{-2tc R }}{Rt^2 \sqrt{t^2-1}}\right ) - \frac{1}{R} \right],
\end{eqnarray}
where $A(Z)=0.074+0.35Z \alpha_e $, $B(Z,r)=[1.071-1.97((Z-80) \alpha_e )^2 -2.128 ((Z-80) \alpha_e )^3+0.169 ((Z-80) \alpha_e )^4 ]cr/(cr+0.07(Z \alpha_e )^2 )$
and $R= | \textbf{r} - \textbf{x}|$.

\begin{table}[t]
\caption{Field shift constants ($F$) in MHz/fm$^2$ of the first five low-lying states of Ca$^+$ from a number of relativistic many-body
methods approximated at different levels. Methods from the top to bottom sequence include more physical effects due to inclusion of more electron
correlation effects systematically. Corrections from the Breit and lowest-order QED corrections are quoted separately towards the end.
Our final values are given as the results from the CCSDpT method along with the higher-order relativistic corrections.
Results from other calculations are given for comparison. Ratios of the differences between results from different states are given in the lower
part of the table and compared with the other calculations and experimental values.}
\begin{ruledtabular}
 \begin{tabular}{l c c c c c}
Method & $4s \ ^2S_{1/2}$ & $4p \ ^2P_{1/2}$ & $4p \ ^2P_{3/2}$ & $3d \ ^2D_{3/2}$ & $3d \ ^2D_{5/2}$   \\
\hline
&&\\
DHF      & $-214.537$ & $-0.413$ & $\sim 0.0$  &  $\sim 0.0$  &  $\sim 0.0$     \\
MBPT(2)  & $-253.635$ &17.061 &17.412 &77.605 &77.170      \\
MBPT(3)  &$-249.506$  &21.130 &21.459 &91.401 &90.767\\
LCCSD    & $-273.053$ &21.966 &22.377 &116.783&116.835     \\
CCSD$^{(2)}$ &$-262.486$&$20.579$&$20.951$&$100.208$&$100.114$ \\
CCSD     & $-259.492$ &20.715 &20.937 &103.269&102.698      \\
CCSDpT   &$-252.604$&25.619&25.838&115.929&115.538\\
\hline
& & \\
$+$ Breit       &0.343& $-0.031$ & $-0.031$ & $-0.006$ &0.057   \\
 $+$QED         & $-0.781$ &0.077&0.069&0.339&0.341   \\
\hline
 & & \\
Total           & $-253.042$ &25.665 & 25.876 & 116.262 & 115.936 \\
\hline
  & & \\
Others \cite{saf1}     &  $-266.6$ & 19.6 & 19.9  & 111.8 & 111.2   \\
\end{tabular}

\begin{tabular}{l c c c}
Method & $\frac{ 4s \ ^2S_{1/2} - 4p \ ^2P_{3/2} } { 4s \ ^2S_{1/2} - 4p \ ^2P_{1/2} } $ & $\frac{3d \ ^2D_{3/2} - 4p \ ^2P_{1/2} }
{ 4s \ ^2S_{1/2} - 4p \ ^2P_{1/2} } $ & $\frac{3d \ ^2D_{3/2} - 4p \ ^2P_{1/2} } { 4s \ ^2S_{1/2} - 4p \ ^2P_{3/2} }$ \\
\hline
&&\\
DHF      & 1.0019 & $-0.0019$  &  $-0.0019$  \\
MBPT(2)  & 1.0013 &  $-0.2237$ & $-0.2234$ \\
MBPT(3)  & 0.9988   &  $-0.2597$ & $-0.2593$ \\
LCCSD    & 1.0014 & $-0.3214$ & $-0.3209$ \\
CCSD$^{(2)}$ & 1.0013   &  $-0.2813$  & $-0.2809$ \\
CCSD     &  1.0008 & $-0.2946$ & $-0.2944$ \\
CCSDpT   &  1.0008 & $-0.3246$ & $-0.3243$ \\
Final    &  1.0007 & $-0.3251$ & $-0.3248$ \\
& &  & \\
Ref. \cite{saf1} & 1.0010 & $-0.3222$ & $-0.3218$ \\
Ref. \cite{berengut} & 1.0 &  &  \\
Expt \cite{shi} & 1.0085(12) & $-0.3114(10)$ & $-0.3088(10)$ \\
\end{tabular}

\end{ruledtabular}
\label{fscon}
\end{table}

We formulate wave functions by treating the considered states of Ca$^+$ as a closed-core $[3p^6]$ of Ca$^{2+}$ with a valence orbital from
different orbital angular momentum for the computational convenience. Again to demonstrate propagation of electron correlation effects
from lower to all-order many-body methods systematically, we adopt Bloch's prescription and express atomic wave function of a state
$\vert \Psi_v \rangle$ with the closed-core $[3p^6]$ and a valence orbital $v$ as \cite{lindgren}
\begin{eqnarray}
 \vert \Psi_v \rangle = \Omega_v \vert \Phi_v \rangle,
\end{eqnarray}
where $\Omega_v$ and $\vert \Phi_v \rangle$ are referred to as the wave operator and the DHF wave function, respectively.
The DHF wave function is constructed as $\vert \Phi_v \rangle= a_v^{\dagger} \vert \Phi_0 \rangle$ for the respective state with the
valence orbital $v$ and the DHF wave function of the closed-core  $\vert \Phi_0 \rangle$. To generate the DHF orbitals, we use Gaussian type
orbital (GTO) basis functions, which are defined as 
\begin{eqnarray}
| \phi^l(r) \rangle &=& r^l \sum_{\nu =1}^{N_l} c_{\nu}^l e^{-\alpha_{\nu} r^2} |\chi(\theta, \varphi) \rangle,
\label{anbas}
\end{eqnarray}
for a orbital with orbital angular momentum $l$, where $|\chi(\theta, \varphi)\rangle $ represents for the angular momentum part, $N_l$ corresponds to 
the total number of analytic functions considered in the calculations and $\alpha_{\nu}$ is an arbitrary coefficient 
constructed satisfying the even tempering condition between two parameters $\alpha_0$ and $\beta$ as 
\begin{eqnarray}
\alpha_{\nu} &=& \alpha_0 \beta^{\nu-1}.
\label{evtm}
\end{eqnarray}
We have chosen 40 GTOs per each $l$ value, $\alpha_0=0.00715$ and $\beta=1.92$ in this work. 

In our approach, the electron correlation effects from different electrons are included by dividing $\Omega_v$ as
\begin{eqnarray}
 \Omega_v =  1+ \chi_0  + \chi_v ,
\end{eqnarray}
where $\chi_0$ and $\chi_v$ are responsible for carrying out excitations from $\vert \Phi_0 \rangle$ and $\vert \Phi_v \rangle$,
respectively, to account for the electron correlation effects because of the residual interaction $V_{es}=H-H_0$ for the DHF
Hamiltonian $H_0$. The core orbital relaxations are then estimated by operating $\chi_0$ on $\vert \Phi_v \rangle$. In the MBPT method,
we express as
\begin{eqnarray}
 \chi_0 = \sum_k \chi_0^{(k)} \ \ \text{and} \ \ \chi_v=\sum_k \chi_v^{(k)},
\end{eqnarray}
where the superscript $k$ refers to number of $V_{es}$ considered and when it is truncated considering up to $n$ number of $V_{es}$, we
refer to this as MBPT(n) method. Following Bloch's approach, the amplitudes of the $\chi_0$ and $\chi_v$ operators are obtained by \cite{lindgren}
\begin{eqnarray}
 [\chi_0^{(k)},H_0]P_0 &=& Q_0 V_{es}(1+ \chi_0^{(k-1)} )P_0
\end{eqnarray}
and
\begin{eqnarray}
[\chi_v^{(k)},H_0]P_v &=& Q_v V_{es} (1+ \chi_0^{(k-1)}+ \chi_v^{(k-1)}) P_v \nonumber \\
 && -  \sum_{m=1 }^{k-1}\chi_v^{(k-m)}  P_v V_{es} \nonumber \\ && \times (1+\chi_0^{(m-1)}+\chi_v^{(m-1)})P_v,
 \label{mbsv}
\end{eqnarray}
where the projection operators are defined as $P_0=\vert \Phi_0 \rangle \langle \Phi_0 \vert $ and $P_v=\vert \Phi_v \rangle \langle
\Phi_v \vert$. The corresponding orthogonal space operators are given by $Q_0 = 1- P_0$ and $Q_v=1-P_v$. Clearly, $\chi_0$ and $\chi_v$ are in normal 
order form with respect to $\vert \Phi_0 \rangle$. We use the normal ordering operators with respect to $\vert \Phi_0 \rangle$ for calculating different 
properties of Ca$^+$.

\begin{figure}[t]
\begin{center}
\includegraphics[width=8.5cm,height=6.0cm]{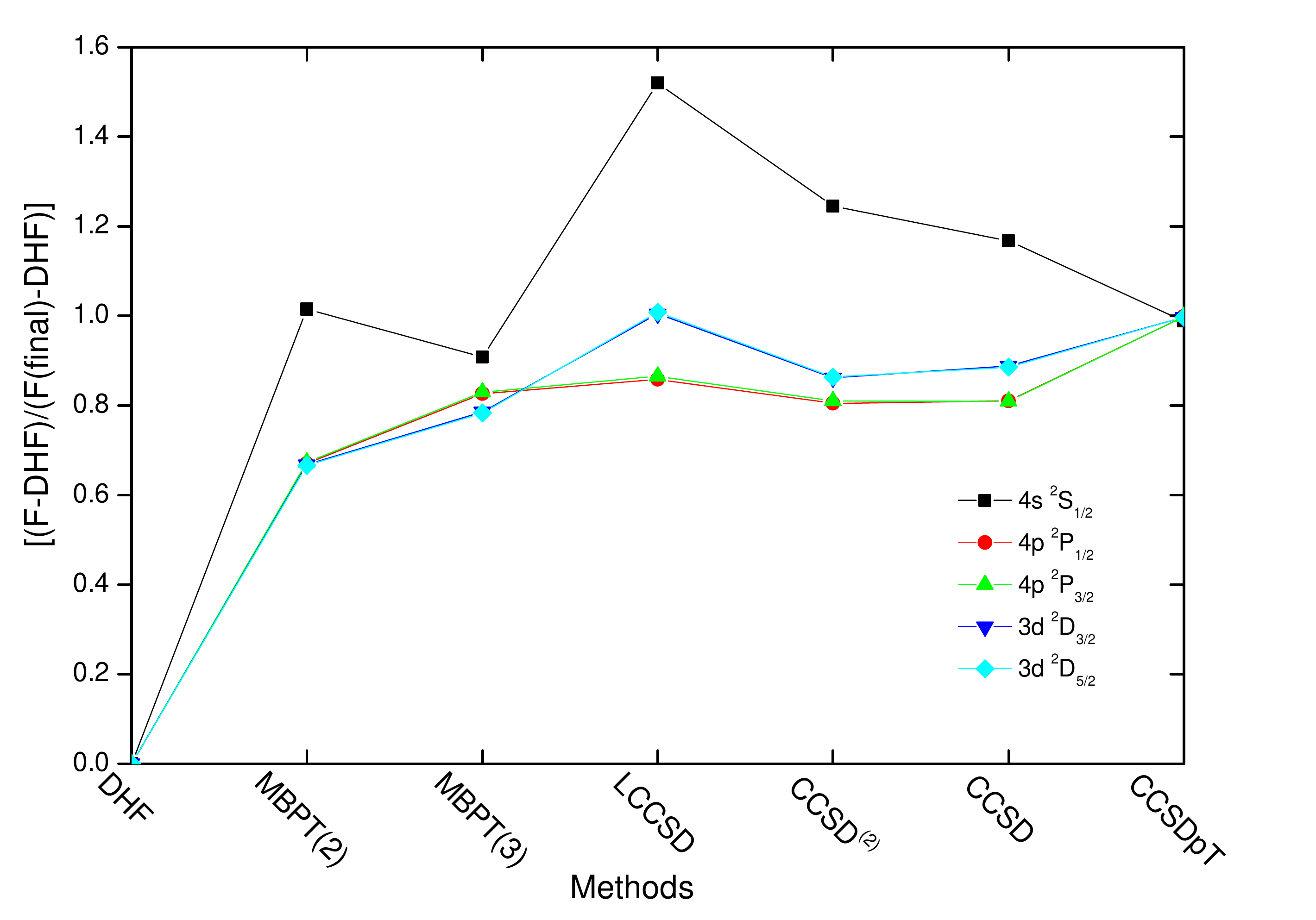}
\end{center}
\caption{(Color online) Demonstration of propagation of electron correlation effects in the evaluation of the field shift constants from lower- to 
all-order methods with different levels of approximations. Methods are quoted in the X-axis with arbitrary unit for the representation purpose only.}
\label{fig1}
\end{figure}

 The electron attachment energy of a state $\vert \Psi_v \rangle$ is evaluated by using an effective Hamiltonian
\begin{eqnarray}
 H_v^{eff}= P_v H_N \Omega_v P_v,
 \label{efhm}
\end{eqnarray}
where the normal order Hamiltonian $H_N= H - P_0 HP_0$ is used. In the MBPT(2) method, we evaluate attachment energy using the expression
$H_v^{eff}= \sum_{k=0}^{1} P_v H_N \Omega_v^{(k)} P_v$.

\begin{sidewaystable}
\caption{Results obtained for $A_{hf}$ (in MHz) and $g_{J}$ factors for the considered states in $^{43}$Ca$^+$ using a number of relativistic many-body methods.
Corrections from the Breit ad QED interactions are quoted separately from the CCSD method. The final results from the CCSDpT method and the
higher order relativistic corrections are compared with the precisely available experimental values. Other calculations on $A_{hf}$ in
$^{43}$Ca$^+$ are summarized in Ref. \cite{ca+bks1} and other theoretical result on $g_J$ factor is given here.}
\begin{ruledtabular}
\begin{tabular}{lcccccccccc}
Methods  & \multicolumn{2}{c}{$4s \ ^2S_{1/2}$} & \multicolumn{2}{c}{$4p \ ^2P_{1/2}$} & \multicolumn{2}{c}{$4p \ ^2P_{3/2}$} & \multicolumn{2}{c}{$3d \ ^2D_{3/2}$} & \multicolumn{2}{c}{$3d \ ^2D_{5/2}$}  \\
\cline{2-3}  \cline{4-5}  \cline{6-7} \cline{8-9}  \cline{10-11}
&&\\
  & $A_{hf}$ & $g_{J}$ & $A_{hf}$ & $g_{J}$ &  $A_{hf}$  & $g_{J}$ & $A_{hf}$ & $g_{J}$ & $A_{hf}$ & $g_{J}$ \\
\cline{2-3}  \cline{4-5} \cline{6-7}  \cline{8-9}  \cline{10-11}
&&\\
DHF           & $-587.9$ & $2.002273$ & $-101.5$  & $0.665863$ & $-19.6$ & $1.334082$ & $-33.2$ & $0.799458$ & $-14.2$ & $1.200381$\\
MBPT(2)      & $-787.2$ & $2.001871$ & $-139.5 $  & $0.665684$ & $-29.7$ & $1.333777$ & $-42.2$ & $0.798178$ & $-5.8$  & $1.197680$\\
MBPT(3)      & $-791.8$ & $2.002313$ & $-140.6$   & $0.669027$ & $-30.1$ & 1.334672 & $-42.2$ & 0.781593  & $-5.3$  & 1.186231 \\
LCCSD        & $-852.1$ & $1.999070$ & $-145.3$   & $0.665901$ & $-33.1$ & $1.333605$ & $-49.6$ & $0.800534$ & $-5.9$  & $1.198074$\\
CCSD$^{(2)}$ & $-814.9$ & $1.999499$ & $-146.3$   & $0.666088$ & $-30.5$ & $1.333659$ & $-47.1$ & $0.799704$ & $-4.5$  & $1.167333$\\
CCSD         & $-810.0$ & $2.002703$ & $-144.5$   & $0.665796$ & $-30.7$ & $1.334156$ & $-47.1$ & $0.799003$ & $-3.9$  & $1.200896$\\
CCSDpT       & $-809.6$ & $2.001995$ & $-145.0$   & $0.665520$ & $-30.9$ & $1.333723$ & $-47.4$ & $0.800041$ & $-4.3$  & $1.200343$\\
\hline
\multicolumn{11}{c}{×}\\
$+$Breit     & $-0.62$  & $-0.0000032$ & $-0.04$ & $0.0000042$ & $-0.03$ &$0.0000013$  &$-0.13$ & $0.0000384$ & $-0.04$ & $0.0000081$\\
$+$QED       & $2.78$   & $-0.0000003$ & $0.002$ & $0.0000003$ & $0.01$  &$-0.0000001$ &$0.04$  & $0.0000002$ & $0.04$  & $-0.0000001$\\
$\Delta g^{Q}_J$ &          & $0.002321$   &         & $-0.000773$ &         &$0.000773$   &        &  $-0.000465$&         & $0.000465$  \\
\hline
&&\\
Total        & $-807.4$ & $2.002267$ & $-145.0$ & $0.665636$ & $-30.9$ & $1.333861$ & $-47.5$ & $0.798554$ & $-4.3$ & $1.200341$\\
\hline
 Others \cite{tommaseo}    &   &   2.0022564(6)(12)    &                                &                             &  &       \\
Experiment &$-806.40207160(8)$ \cite{arbes} & 2.00225664(9) \cite{tommaseo} & $-145.4(1)$ \cite{norterhauser} & &$-31.0(2)$ \cite{norterhauser} & &$-47.3(2)$ \cite{norterhauser} & &$-3.8931(2)$ & 1.2003340(25) \cite{chwalla} \\
\end{tabular}
\end{ruledtabular}
\label{ahfgj}
\end{sidewaystable}

To analyze the correlation effects in the methods to all-order in $V_{es}$, we employ the RCC theory in the Fock-space formalism using the
exponential {\it ansatz} of the wave function as
\begin{eqnarray}
 \vert \Psi_v \rangle  \equiv  \Omega_v \vert \Phi_v \rangle &=& e^{ \{ T + S_v \} } \vert \Phi_v \rangle \nonumber \\
 &=& e^T \{ 1+ S_v \} \vert \Phi_v \rangle,
 \label{eqcc}
\end{eqnarray}
which implies that $\chi_0= e^T-1$ and $\chi_v=e^TS_v -1$ with $T$ and $S_v$ are the RCC excitation operators due to $V_{es}$ that excite electrons from
the core and core along with the valence orbitals to the virtual space respectively. It to be noted that expansion of the exponential
form of $S_v$ terminates naturally at the linear level owing to presence of only one valence orbital in the reference DHF state.

We consider only the single and double excitations, denoted by the subscripts $1$ and $2$ respectively, in the CCSD method by
expressing
\begin{eqnarray}
 T=T_1 +T_2 \ \ \ \text{and} \ \ \ S_v = S_{1v} + S_{2v}.
 \label{eqsd}
\end{eqnarray}
This approximation is considered to be good enough to account for electron correlation effects quite reliably to produce many results
in Ca$^+$. Some of the previous calculations \cite{pruttivarasin,safronova,dzuba1,beloy} are reported assuming only the linear terms in
Eq. (\ref{eqcc}) of the CCSD method, referred as LCCSD method, owing to very expensive computations involved with the non-linear terms. For 
representative analysis of correlation trends in this work from the LCCSD method, we approximate Eq. (\ref{eqcc}) to
\begin{eqnarray}
 \vert \Psi_v \rangle &=& \{ 1+ T + S_v \} \vert \Phi_v \rangle .
 \label{eqlcc}
\end{eqnarray}

The amplitudes of the RCC operators are evaluated using the equations
\begin{eqnarray}
 \langle \Phi_0^* \vert H_N + H_NT   \vert \Phi_0 \rangle &=& 0
\label{eqlt}
 \end{eqnarray}
and
\begin{eqnarray}
 \langle \Phi_v^* \vert \big ( H_N - \Delta E_v \big ) S_v \vert \Phi_v \rangle &=&  - \langle \Phi_v^* \vert \big ( H_N + H_NT \big ) \vert \Phi_v \rangle ,\nonumber \\
\label{eqlsv}
 \end{eqnarray}
in the LCCSD method approximation. Similarly, these amplitudes are obtained in the CCSD method approximation by the solving the following
equations
\begin{eqnarray}
 \langle \Phi_0^* \vert \overline{H}_N  \vert \Phi_0 \rangle &=& 0
\label{eqt}
 \end{eqnarray}
and
\begin{eqnarray}
 \langle \Phi_v^* \vert \big ( \overline{H}_N - \Delta E_v \big ) S_v \vert \Phi_v \rangle &=&  - \langle \Phi_v^* \vert \overline{H}_N \vert \Phi_v \rangle .
\label{eqsv}
 \end{eqnarray}
In these expressions $\vert \Phi_0^* \rangle$ and $\vert \Phi_v^* \rangle$ are the excited state configurations, here up to doubles,
with respect to the DHF states $\vert \Phi_0 \rangle$ and $\vert \Phi_v \rangle$ respectively and $\overline{H}_N= \big ( H_N e^T \big )_l$
with subscript $l$ representing for the linked terms only. The attachment energy $\Delta E_v$ of the electron in the valence orbital
$v$ following Eq. (\ref{efhm}) is evaluated as
\begin{eqnarray}
 \Delta E_v  = \langle \Phi_v \vert \left \{ H_N + H_NT + H_NS_v \right \} \vert \Phi_v \rangle
 \label{eqleng}
\end{eqnarray}
in the LCCSD method and
\begin{eqnarray}
 \Delta E_v  = \langle \Phi_v \vert \overline{H}_N \left \{ 1+S_v \right \} \vert \Phi_v \rangle
 \label{eqeng}
\end{eqnarray}
in the CCSD method framework.

  After obtaining amplitudes of the wave operators using the equations described earlier, the considered properties are evaluated as the
the expectation values of the respective operators. Representing them by a general operator $O$, the expectation value in the state
$\vert \Psi_v \rangle$ is evaluated by
\begin{eqnarray}
\frac{\langle \Psi_v \vert O \vert \Psi_v \rangle}{ \langle \Psi_v \vert \Psi_v \rangle }
&=& \frac {\langle \Phi_v \vert \Omega_v^{\dagger} O_N \Omega_v \vert \Phi_v\rangle}
{\langle \Phi_v \vert \Omega_v^{\dagger} \Omega_v \vert \Phi_v \rangle }  ,
\label{preq}
\end{eqnarray}
with $O_N=O-P_0 O P_0$. This in the MBPT(n) method corresponds to
\begin{widetext}
\begin{eqnarray}
\frac{\langle \Psi_v \vert O \vert \Psi_v \rangle}{ \langle \Psi_v \vert \Psi_v \rangle }
&=& \frac { \sum_{k=0}^{\text{n}} \sum_{m=0}^{\text{n}} \langle \Phi_v \vert [ 1 + \chi_0^{\dagger (m) } + \chi_v^{\dagger (m)} ]
O_N  [ 1+ \chi_0^{(k)} + \chi_v^{(k)} ] \vert \Phi_v\rangle}
{ \sum_{k=0}^{\text{n}} \sum_{m=0}^{\text{n}} \langle \Phi_v \vert  [ 1 + \chi_0^{\dagger (m) } + \chi_v^{\dagger (m)} ] [ 1+ \chi_0^{(k)} + \chi_v^{(k)} ] \vert \Phi_v\rangle }  ,
\label{mpreq}
\end{eqnarray}
\end{widetext}
with the constraint that each term in the expansion of the above expression should have $m+k \le n$ orders of $V_{es}$. We consider terms
belonging to both MBPT(2) and MBPT(3) methods for our analysis.

Similarly, the properties in the LCCSD method approximation are evaluated by
\begin{widetext}
\begin{eqnarray}
\frac{\langle \Psi_v \vert O \vert \Psi_v \rangle}{ \langle \Psi_v \vert \Psi_v \rangle }
&=& \frac{\langle \Phi_v | \{O_N + O_NT+ T^{\dagger} O_N + O_N S_v + S_v^{\dagger} O_N + T^{\dagger}O_N T+ S_v^{\dagger}O_NT+ T^{\dagger} O_N S_v + S_v^{\dagger}O_N S_v \} | \Phi_v \rangle}
{\langle \Phi_v \vert \{ 1+ T^{\dagger} T+ S_v^{\dagger} T + T^{\dagger} S_v + S_v^{\dagger} S_v \} \vert \Phi_v \rangle}
\label{lprpeq}
\end{eqnarray}
\end{widetext}
and evaluating the following expression in the CCSD method
\begin{eqnarray}
\frac{\langle \Psi_v \vert O \vert \Psi_v \rangle}{ \langle \Psi_v \vert \Psi_v \rangle }
&=& \frac{\langle \Phi_v \vert \{1+ S_v^{\dagger}\} e^{T^{\dagger}} O_N e^T \{1+S_v\} \vert \Phi_v \rangle} {\langle \Phi_v \vert \{1+ S_v^{\dagger}\}
e^{T^{\dagger}} e^T \{1+S_v \} \vert \Phi_v \rangle }.
\label{prpeq}
\end{eqnarray}

\begin{figure}[t]
\begin{center}
\includegraphics[width=8.5cm,height=6.0cm]{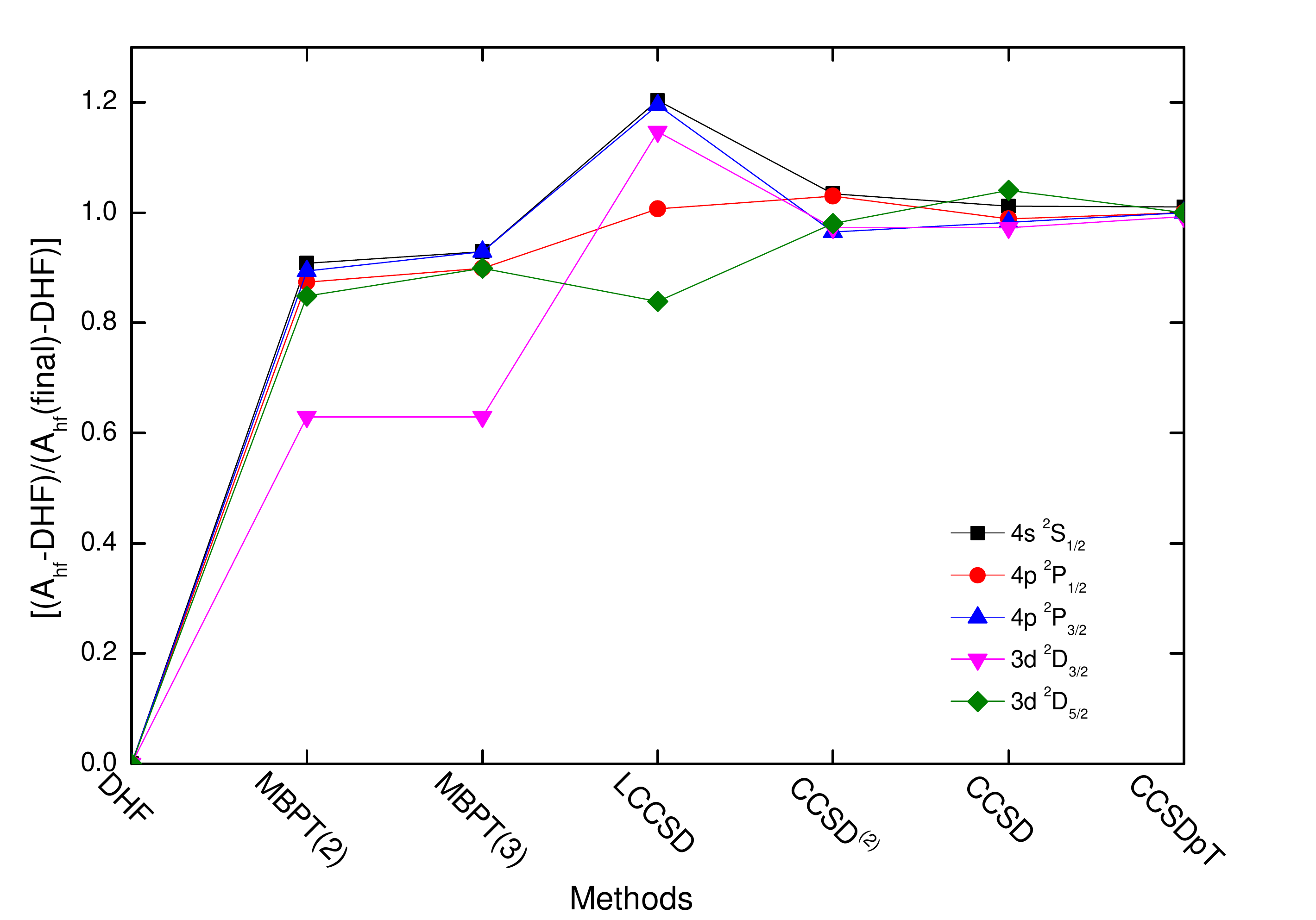}
\end{center}
\caption{(Color online) Correlation trends in the $A_{hf}$ values from lower- to all-order many-body methods with different levels of approximations.
 Methods quoted in the X-axis are with arbitrary unit.}
\label{fig2}
\end{figure}

Contribution coming from $\langle \Phi_v \vert O_N \vert \Phi_v \rangle = \langle \phi_v \vert o \vert \phi_v \rangle$ for the single particle
wave function of the valence orbital $\vert \phi_v \rangle$ and expressing $O=\sum_i o (r_i)$ in the above expressions is termed as the DHF
result here. As can be seen, the property evaluating expressions in the MBPT(2) and LCCSD methods have finite terms while the expression
of the CCSD method has infinite numbers of terms due to non-truncative nature of $e^{T^{\dagger}} O e^T$ and $e^{T^{\dagger}} e^T$. However,
we adopt an iterative procedure to incorporate as much contributions from these non-truncative series as described in our previous works
\cite{nandy1,nandy2}. Further to highlight the differences in the results from the LCCSD and CCSD methods due to the non-linear terms
appearing in the amplitude and property determining expressions, we also evaluate contributions by using the expression of Eq. (\ref{lprpeq})
but substituting amplitudes of the CCSD operators. This approximation is referred to as the CCSD$^{(2)}$ method, which involves non-linear
contributions only through the wave functions.

Uncertainties in our calculations would come from three different sources: (i) consideration of finite size of basis functions in the calculations,
(ii) higher level excitations over the CCSD method and (iii) numerical calculations of various operations. As mentioned earlier we are not focusing
here to give very accurate results, but our intention is to draw attentions on the trends of correlation effects in the evaluation of properties
having different radial and angular behaviors. In this point of view, we do not focus on the basis size truncation error and numerical instabilities. 
Such analyses are already carried out in our previous studies in most of the considered properties \cite{shi,bks1,ca+bks1,ca+bks2,pradeep}. However, 
we would like to demonstrate importance of the higher level excitations contributions that are neglected in the CCSD method approximations in the  
determination of various properties in Ca$^+$. To answer this to some extent, we estimate contributions from the leading order triple excitations that 
are neglected in the CCSD method. For the same, we define perturbative RCC operators to include leading order triple excitations as
\begin{eqnarray}
 T_{3}^{pert}  &=& \frac{1}{6} \sum_{abc,pqr} \frac{\big ( H_N T_2 \big )_{abc}^{pqr}}{\epsilon_a + \epsilon_b + \epsilon_c - \epsilon_p -\epsilon_q - \epsilon_r}
\label{t3eq}
 \end{eqnarray}
 and
\begin{eqnarray}
 S_{3v}^{pert} &=& \frac{1}{4} \sum_{ab,pqr} \frac{\big ( H_N T_2 + H_N S_{2v} \big )_{abv}^{pqr}}{\Delta E_v + \epsilon_a + \epsilon_b - \epsilon_p -\epsilon_q - \epsilon_r} ,
\label{s3eq}
 \end{eqnarray}
as part of the $T$ and $S_v$ operators, respectively. In these definitions, $\{a,b,c \}$ and $\{ p,q,r \}$ represent for the occupied and virtual
orbitals, respectively, and $\epsilon$ are their single particle orbital energies. Contributions from the $T_{3}^{pert}$ and $S_{3v}^{pert}$
operators are estimated using Eq. (\ref{prpeq}). This follows evaluation of extra terms as $T_2^{\dagger}OT_{3}^{pert}$, $T_2^{\dagger}OS_{3v}^{pert}$,
$S_{2v}^{\dagger}OS_{3v}^{pert}$, $S_{1v}^{\dagger}T_2^{\dagger}OS_{3v}^{pert}$, $T_{3}^{pert \dagger}OT_{3}^{pert}$, $S_{3v}^{pert \dagger}OS_{3v}^{pert}$,
and their complex conjugate (c.c.) terms, which are computationally very expensive. Nevertheless, we have put lot of efforts to estimate
contributions from these terms by computing more than 500 diagrams adopting intermediate steps. These diagrams involves at least $n_p^4 n_h^3$,
$n_p^3 n_h^5$, $n_p^5 n_h^3$, $n_p^4 n_h^4$, or $n_p^6 n_h^2$ times allowed multipoles of the RCC operators, with $n_p$ number of virtual orbitals and
$n_h$ number of occupied orbits, even after adopting the intermediate steps of computations. The CCSD values along with these contributions are referred to as the
results from the singles, doubles and partial triples coupled-cluster (CCSDpT) method.

\begin{figure}[t]
\begin{center}
\includegraphics[width=8.5cm,height=6.0cm]{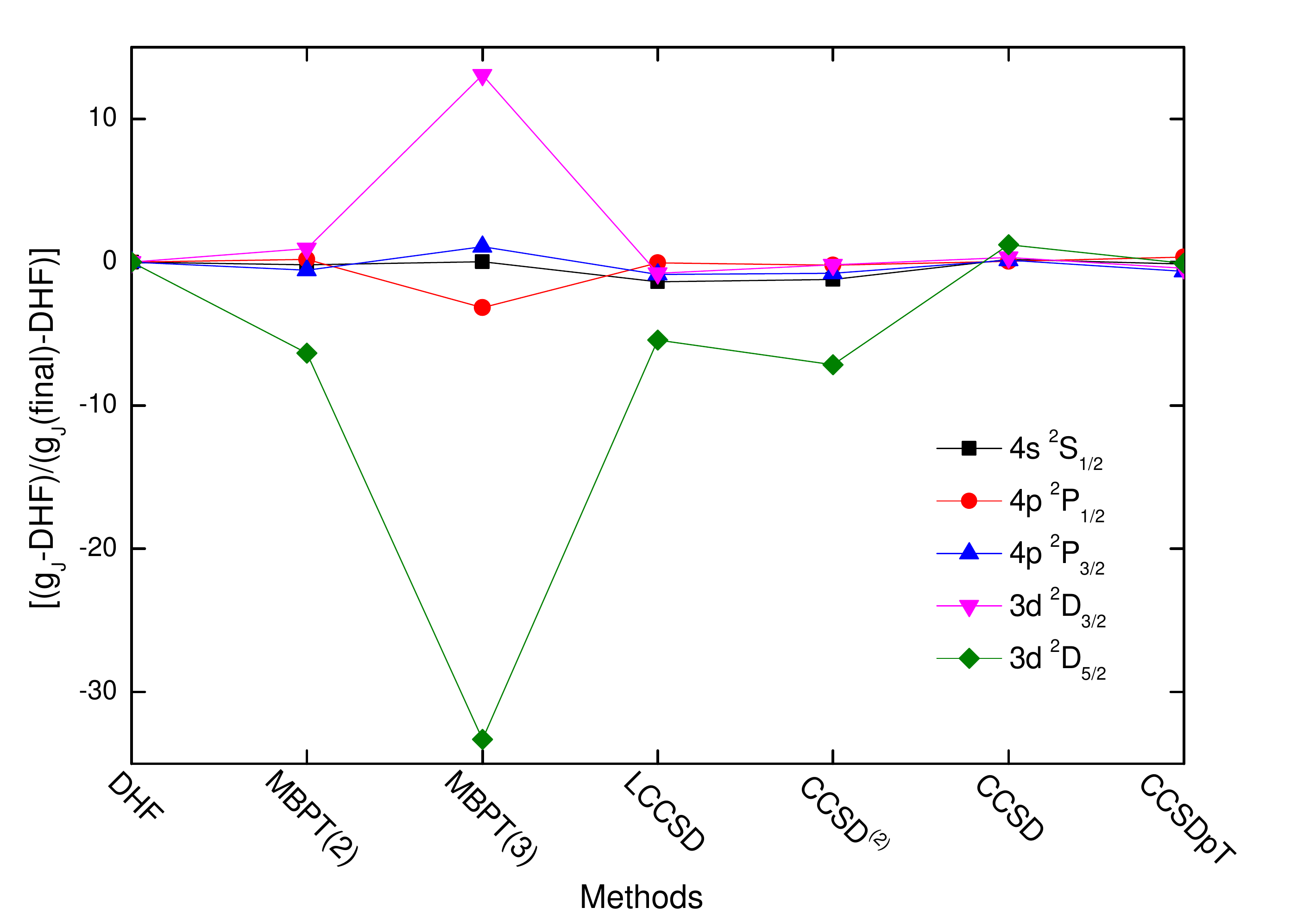}
\end{center}
\caption{(Color online) Correlation trends in the $g_J$ factors from lower- to all-order many-body methods with different levels of approximations.
 Methods quoted in the X-axis are with arbitrary unit.}
\label{fig3}
\end{figure}

\begin{table*}[t]
\caption{Results of B$_{hf}$ (in MHz) and $\Theta$ (in a.u.) from different methods and comparison with other calculations and high-precision experimental 
values. Other theoretical works on $B_{hf}$ are discussed elsewhere \cite{ca+bks1}.}
\begin{ruledtabular}
\begin{tabular}{lccccccc}
Methods &\multicolumn{2}{c}{$4p \ ^2P_{3/2}$} & \multicolumn{2}{c}{$3d \ ^2D_{3/2}$}& \multicolumn{2}{c}{$3d \ ^2D_{5/2}$}&    \\
\cline{2-3}  \cline{4-5}  \cline{6-7}
&&\\
    &$B_{hf}$&$\Theta$&$B_{hf}$&$\Theta$&$B_{hf}$&$\Theta$ \\
\cline{2-3}  \cline{4-5}  \cline{6-7}
&&\\
DHF          & $-4.301$ & $4.670$ & $-2.417$ & $1.712$ & $-3.425$ & $2.451$\\
MBPT(2)     & $-6.420$ & $4.334$ & $-3.030$ & $1.266$ & $-4.295$ & $1.814$\\
MBPT(3)     & $-6.462$ & $4.332$ & $-2.974$ & $1.265$ & $-4.147$ & $1.812$\\
LCCSD       & $-7.091$ & $4.279$ & $-3.104$ & $1.208$ & $-4.399$ & $1.732$\\
CCSD$^{(2)}$& $-6.711$ & $4.339$ & $-3.160$ & $1.287$ & $-4.443$ & $1.844$\\
CCSD        & $-6.700$ & $4.341$ & $-3.030$ & $1.291$ & $-4.235$ & $1.849$\\
CCSDpT      & $-6.702$ & $4.338$ & $-3.027$ & $1.289$ & $-4.117$ & $1.846$\\
\hline
\multicolumn{7}{c}{×}\\
$+$Breit   & $0.0079$  & $-0.0006$  & $-0.0006$ & $-0.002$ & $-0.0041$ &$-0.002$\\
$+$QED     & $-0.0003$ & $\sim 0.0$ & $-0.0002$ & $0.0$   & $-0.0003$ &$0.0$\\
\hline
\multicolumn{7}{c}{×}\\
Total&$-6.694$&$4.337$&$-3.028$&$1.287$&$-4.121$&$1.844$\\
\hline
\multicolumn{7}{c}{×}\\
Ref. \cite{jiang}           &                           &       &  &   1.289(11)     &    &   1.849(17) \\
Ref. \cite{ca+bks2}          &                           &       &  &   1.338         &    &   1.917 \\
Experiment &$-6.7(1.4)$ \cite{silverans}; $-4.2(1.3)$ \cite{ruiz} & &$-3.7(1.9)$ \cite{norterhauser} & &$-4.241(4)$ \cite{benhelm} &$1.83(1)$\\
\end{tabular}
\end{ruledtabular}
\label{bhfth}
\end{table*}

\section{Results and Discussion}

Before discussing the trends of electron correlation effects in Ca$^+$ for evaluating the considered properties, we give results of these quantities
obtained from various approximated many-body methods and compare them against the available experimental values. First, we compare results in 
many of the low-lying states from various approximated methods and strive to perceive their underlying trends in the electron correlation contributions 
to determine the spectroscopic properties. This is done for the individual properties and compared with the available experimental results and other theoretical 
calculations. We proceed first with the results for the field shift constants in which the associated operator is a scalar.  This follows with the 
discussions about the $A_{hf}$ and $g_J$ values, that are described by the operators having the same rank (i.e. one) but different radial behaviors.
Then, we present the $B_{hf}$ and $\Theta$ values that are finite only for the states with angular momenta $J > 1/2$ following the triangle condition 
for their physical operators of rank two.

\begin{figure}[t]
\begin{center}
\includegraphics[width=8.5cm,height=6.0cm]{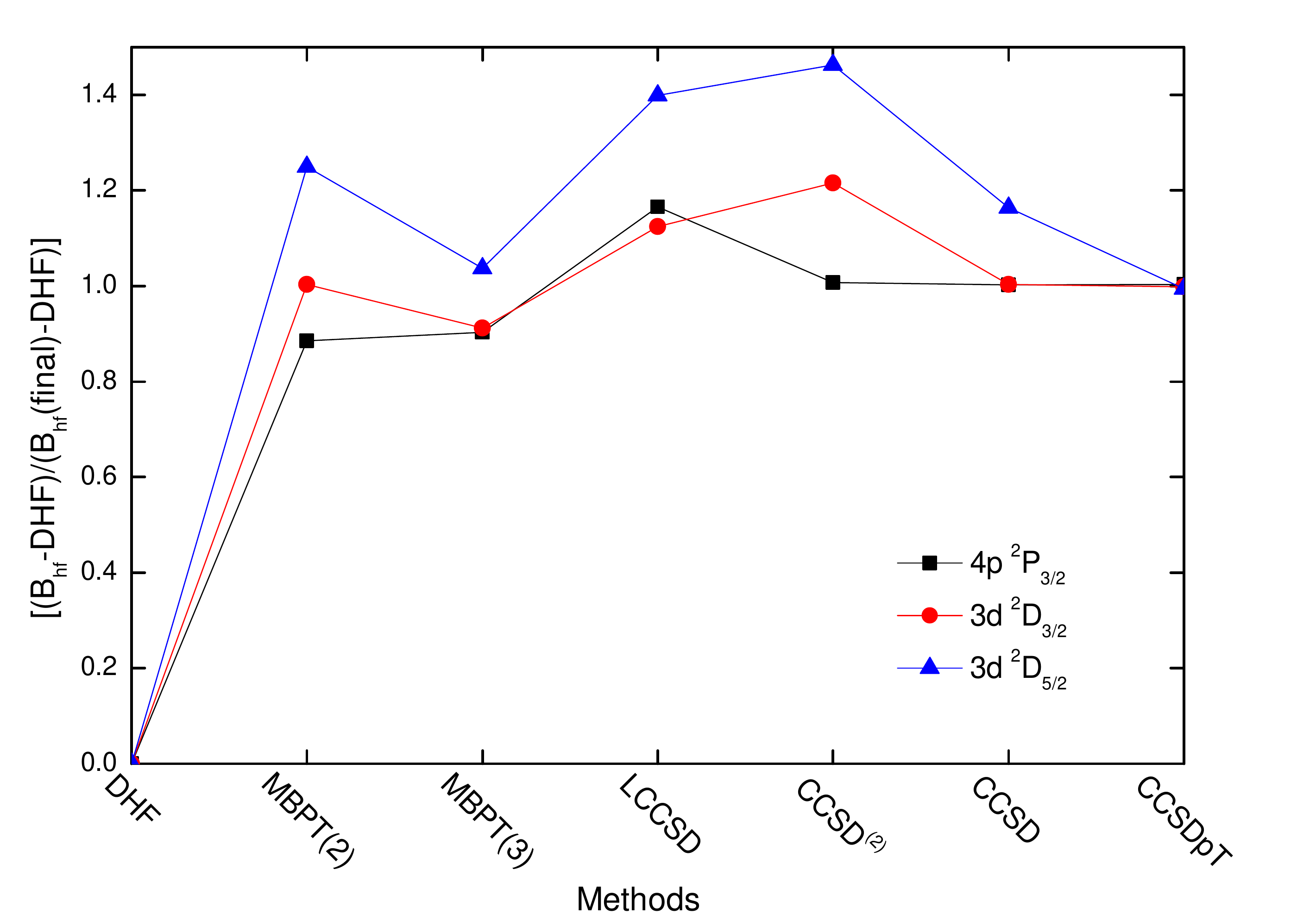}
\end{center}
\caption{(Color online) Correlation trends in the $B_{hf}$ values from lower- to all-order many-body methods with different levels of approximations.
 Methods quoted in the X-axis are with arbitrary unit.}
\label{fig4}
\end{figure}

The field shift constants obtained from different employed methods for the considered states are given in Table \ref{fscon}. These
quantities cannot be measured directly for a state in an experiment, but their differences between any two states can be extracted out from the
isotope shift measurement of the transition involving the states using the King's plot \cite{king}. Our final values are given as the results
obtained using the CCSDpT method along with the relativistic corrections from the Breit and QED interactions. We have compared our results
with another calculations available using the MBPT(3) method \cite{saf1} in Table \ref{fscon}. As seen, results from our MBPT(3) method and
that are given in Ref. \cite{saf1} differ a lot. This is because there are also additional contributions included from the higher-order random-phase
approximation (RPA) with the MBPT(3) contributions in Ref. \cite{saf1}. In fact, one more calculation is available on the field shift
contributions to the D1 and D2 lines of Ca$^+$, but it does not give field shift constants for the individual state \cite{berengut}.

\begin{figure}[t]
\begin{center}
\includegraphics[width=8.5cm,height=6.0cm]{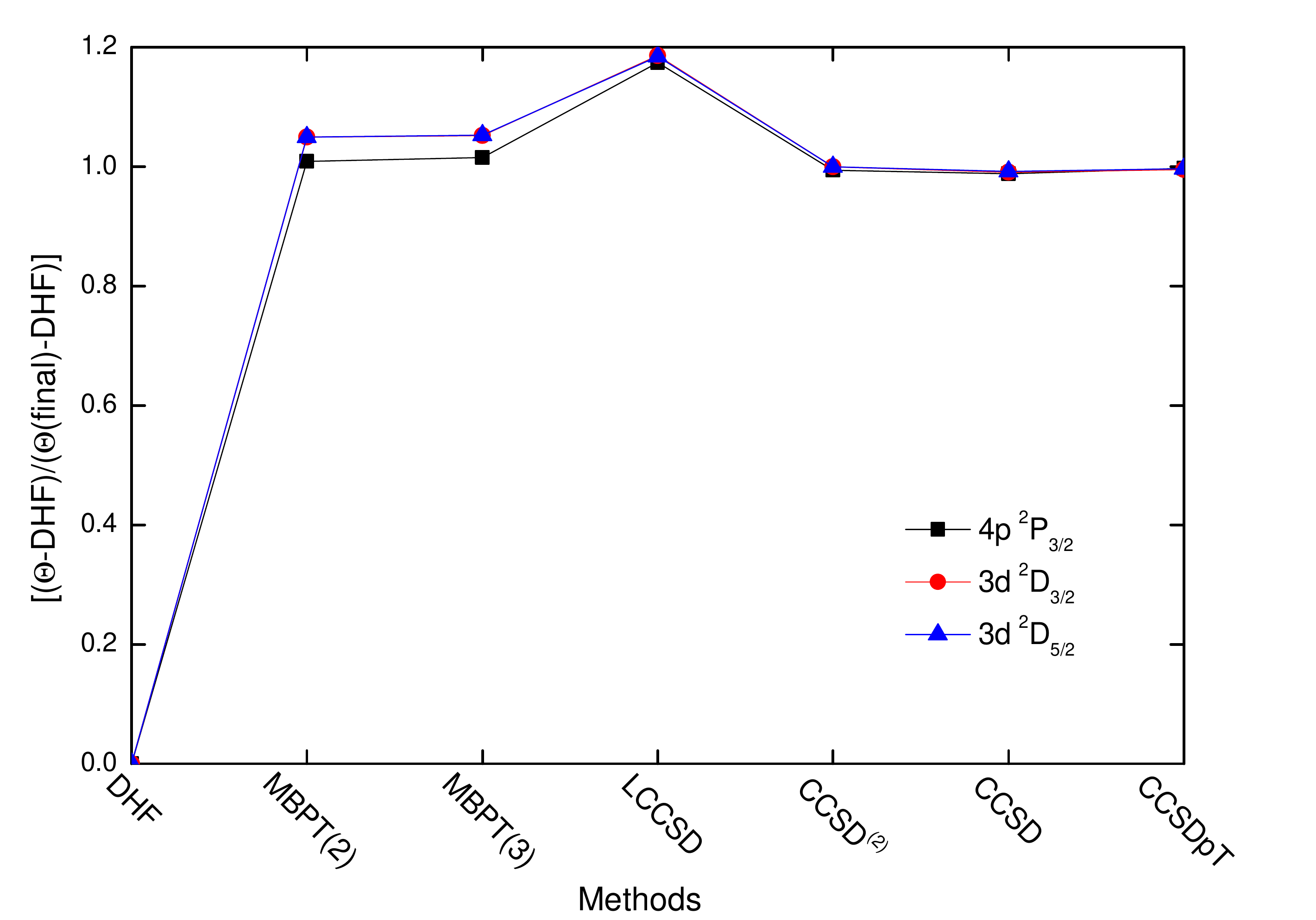}
\end{center}
\caption{(Color online) Correlation trends in the $\Theta$ values from lower- to all-order many-body methods with different levels of approximations.
 Methods quoted in the X-axis are with arbitrary unit.}
\label{fig5}
\end{figure}

The differences in the field shift contributions in Ca$^+$ for the D1, D2 and $3d \ ^2D_{3/2}
\rightarrow 4p \ ^2P_{1/2}$ transitions were extracted out in the recent measurements \cite{shi,gebert}. It is  observed in these measurements that
ratios of the field shifts between the above transitions do not agree with the previously available calculations. To investigate it further
using our results from all the considered methods at different levels of approximation, we present ratios for these transitions and compare them
against the experimental values towards the end of Table \ref{fscon}. As can be seen, the ratio of $F$ values among the D2 and D1
lines do not agree with the experimental values at all. It gives a slightly larger value at the DHF method, but then the ratio remains almost
constant around 1.0010 from the all-order methods. The CCSD and CCSDpT methods give a smaller value about 1.0008 even when the $F$ values
for the individual states differ significantly, and finally it settles down at 1.0007 after considering the relativistic corrections. This
is in close agreement with the value given by a combined configuration (CI) and MBPT method in Ref. \cite{shi}. We had reported this
ratio as 1.0029 from the CCSD method in Ref. \cite{shi}, however when more effective two-body contributions from the non-linear
terms of the truncative series appearing in Eq. (\ref{prpeq}) are included the ratio came down drastically. We had also estimated partial
triples effects by considering many important diagrams representing the $T_2^{\dagger}OS_{3v}^{pert}$ RCC term in Ref. \cite{shi}. We, however,
find contributions from the other aforementioned terms containing the triply excited perturbed operators cancel out most
of the contributions from the above term. The net partial triples contributions are still significant to the field shift constants,
but the ratio between the D2 and D1 lines is not affected much. The results from Ref. \cite{saf1} gives this ratio as 1.0010,
while the other calculation based on the finite gradient approach in the Green function technique offers this ratio as one \cite{berengut}.
In contrast, the ratios among the field shift of the $3d \ ^2D_{3/2} \rightarrow 4p \ ^2P_{1/2}$ transition with the D1 and D2 lines show
very different trends. The DHF method gives low ratios and gradually their values increase and then decrease; finally producing values
towards the experimental results. In fact, large differences in the $F$ values from the CCSDpT and CCSD methods suggest
that the electron correlation effects play crucial roles to determine the $3d \ ^2D_{3/2;5/2}$ states. This also helps in improving the ratios
over the results from the CCSD method. It, however, gives almost the same ratios between the $F$ values of the $3d \ ^2D_{3/2} \rightarrow
4p \ ^2P_{1/2}$ transition with the D1 and D2 lines. Thus, it supports the fact that the $4p \ ^2P_{1/2}$ and $4p \ ^2P_{3/2}$ states behave
differently in the nuclear region which cannot be described by determining the nuclear potential in the uniform charge distribution
approximation.

In Fig. \ref{fig1}, we plot the field shift constants from different methods normalizing with the final values of the respective states. In these
results we have subtracted out the DHF contributions in order to highlight the propagation of the electron correlation effects through all the considered
many-body methods. As seen, the correlation effects in the ground state seem to be very large as its valence $s$-orbital has larger overlap in the
nuclear region. In different states the correlation effects are seen to be enhanced through lower to higher-order many-body methods, except a slight
deviation in the ground state. The MBPT(2) method estimates little larger correlation effects than the MBPT(3) method and the LCCSD method overestimates 
these effects in the ground state. There are also noticeable differences between the results from the CCSD$^{(2)}$ and CCSD methods indicating that the 
non-linear terms from Eq. (\ref{prpeq}) play important roles in determining the $F$ values.

\begin{figure}[t]
\begin{center}
\includegraphics[width=8.5cm,height=6.0cm]{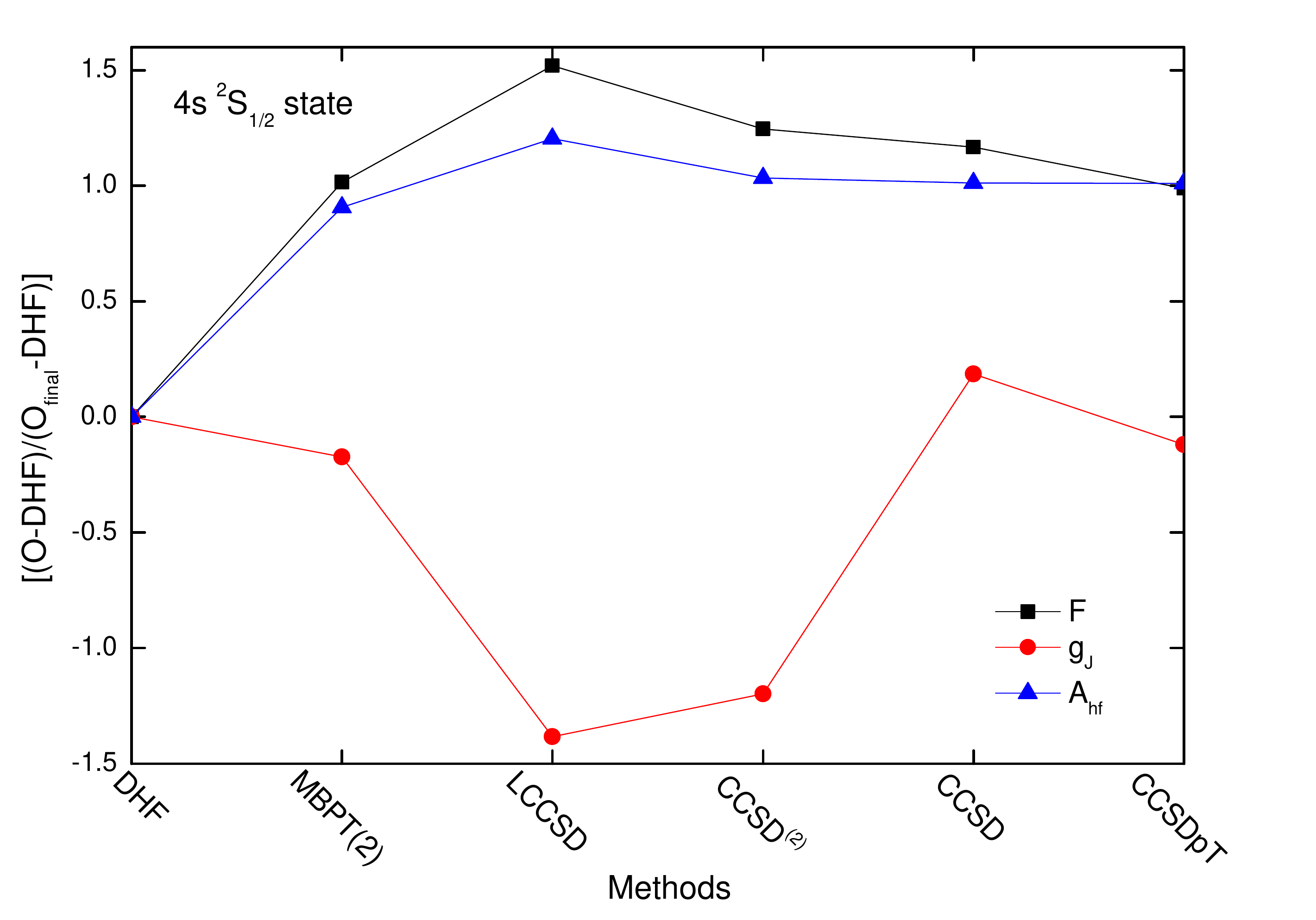}
\end{center}
\caption{(Color online) Comparison of correlation trends among various properties with different radial and angular momentum dependencies through
the employed many-body methods in the ground state of $^{43}$Ca$^+$.}
\label{fig6}
\end{figure}

In Table \ref{ahfgj}, we present the $A_{hf}$ and $g_J$ values together from the employed methods for all the considered states in $^{43}$Ca$^+$.
Available precise experimental results of these quantities are also quoted at the end of the table \cite{tommaseo,chwalla,arbes,norterhauser}. As
mentioned in Sec. \ref{sec2}, the corresponding electronic operators to evaluate these quantities have the same angular factors but with different
radial dependencies. Comparison in the trends of the results from the DHF to CCSDpT methods demonstrate, they do not follow similar pattern in the
propagation of the electron correlation effects. For example, the DHF value of $A_{hf}$ for the $4s ~ ^2S_{1/2}$ state is about two-third of the
experimental value, while the the DHF value of $g_J$ in this state is very close to the experimental value. Inclusion of the correlation effects
through the MBPT(2) method brings the $A_{hf}$ value towards the experimental result in the $4s ~ ^2S_{1/2}$ state, but the LCCSD method
overestimates the value while the CCSD method takes it again very close to the experimental result. Addition of higher-order relativistic corrections
from the Breit and QED interacts improve the results further, so also the triples contributions from the CCSDpT method. This trend is almost
similar in other states except in the $3d ~ ^2D_{5/2}$ state of $^{43}$Ca$^+$ for the evaluation of $A_{hf}$. In the $3d ~ ^2D_{5/2}$ state, the
DHF method predicts a very large value while inclusion of the correlation effects lower the value close to the experimental result. Higher
relativistic corrections and triples contributions through the CCSDpT method are found to be very small. In our previous work \cite{sahoo1}, we
had discussed results from other calculations and also reported values from the CCSD method. However, we had only accounted for linear terms from 
$e^{T^{\dagger}} O_N e^T$ of Eq. (\ref{prpeq}) in that work, while the non-linear terms are included through self-consistent procedure here.

\begin{figure}[t]
\begin{center}
\includegraphics[width=8.5cm,height=6.0cm]{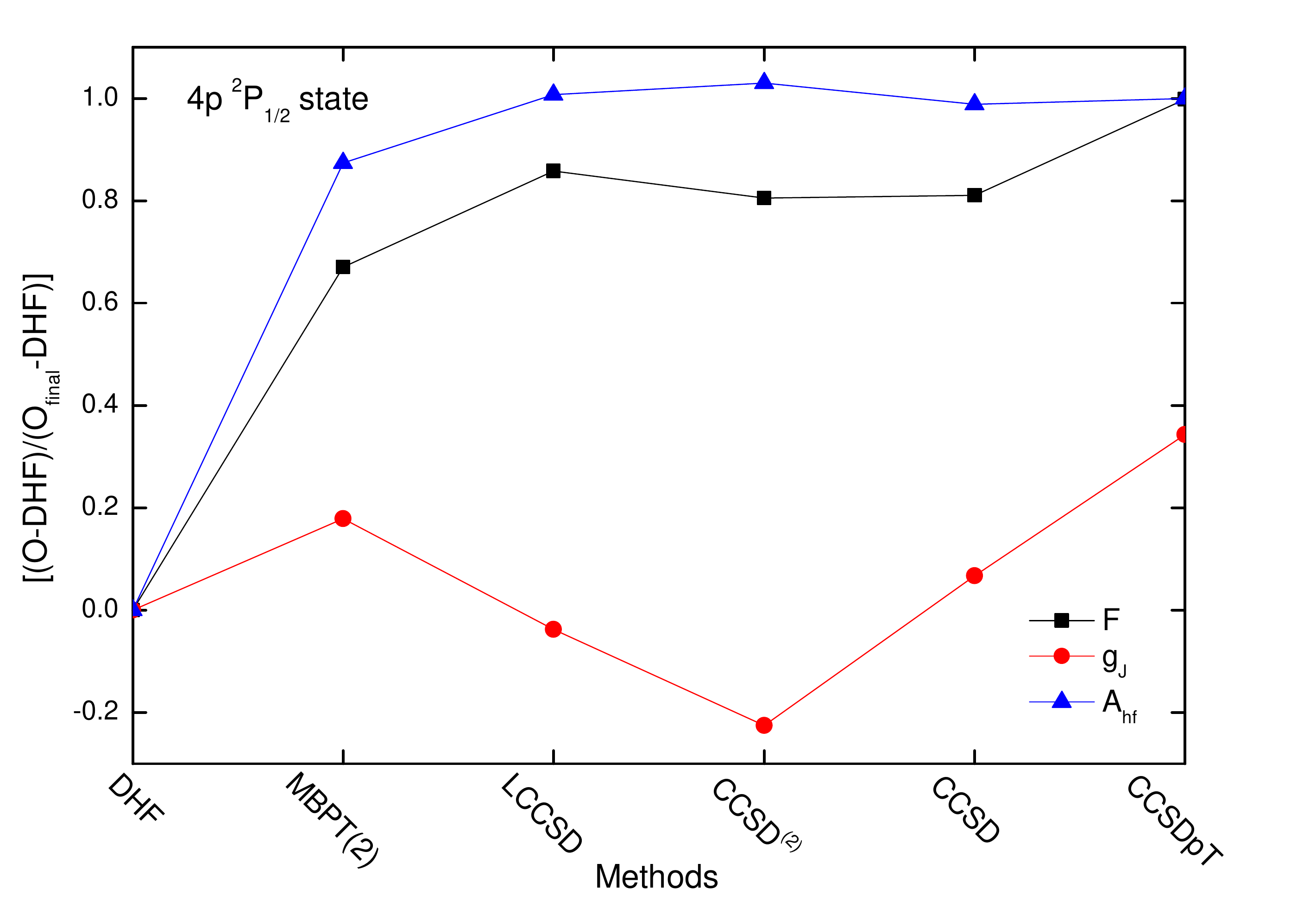}
\end{center}
\caption{(Color online) Comparison of correlation trends among investigated properties in the $4p ~ ^2P_{1/2}$ state of $^{43}$Ca$^+$ through
the employed many-body methods.}
\label{fig7}
\end{figure}

To demonstrate prominently the trends of correlation effects captured through lower to all-order many-body methods, we plot the
($A_{hf}-$DHF)/($A_{hf}({\text{final}})$-DHF) values of $A_{hf}$ from all the employed methods in Fig. \ref{fig2}, where $A_{hf}({\text{final}})$
corresponds to the final result. As can be seen, these trends are different than the $F$ values shown in Fig. \ref{fig1}.
The plots are almost flat from the MBPT(2) to MBPT(3) methods and also from the CCSD to CCSDpT methods implying partial triple effects are negligibly
small. The LCCSD method overestimates the correlation effects in all the states except in the $3d ~ ^2D_{5/2}$ state, in which it underestimates
these effects. In contrast to the $F$ values, we find here there is almost no difference between the results from the CCSD$^{(2)}$ and CCSD
methods. This suggests that the non-linear effects add contributions only through the evaluation of the amplitudes of the wave functions.

Even though $A_{hf}$ and $g_J$ determining expressions have same angular factors, the DHF value of the $g_J$ factor of the ground state is found
to be close to the experimental value in contrast to the $A_{hf}$ value as mentioned above. When the correlation effects are added through the
MBPT(2) and MBPT(3) methods, the results do not improve over the DHF values. In fact, the all-order LCCSD, CCSD$^{(2)}$ and CCSD methods also do not 
give satisfactory results compared with the experimental value \cite{tommaseo}. Corrections from the Breit and QED interactions are found to be extremely 
small with respect to the contributions arising due to the electron correlation effects. The partial triple effects, however, play very important roles 
in achieving result to the closed experimental value. In fact, this is achieved at the cost of huge computational efforts to evaluate contributions from 
more than 500 diagrams. Again, trends of correlation contributions to the excited states are also found to be different than the ground state. In these
states, the triples effects through the CCSDpT method are also coming out to be quite large. Corrections from the Breit and QED interactions are
not so important than the higher-order electron correlation effects. Our result for the  $3d ~ ^2D_{5/2}$ state also agrees reasonably with the
corresponding experimental value \cite{chwalla}. There is only one calculation on the $g_J$ factor available for the ground state using the
MCDF method reported by the same authors who had performed the measurement \cite{tommaseo}. They had used a very larger number of CSFs to
attain the result matching with the experimental value.

\begin{figure}[t]
\begin{center}
\includegraphics[width=8.5cm,height=6.0cm]{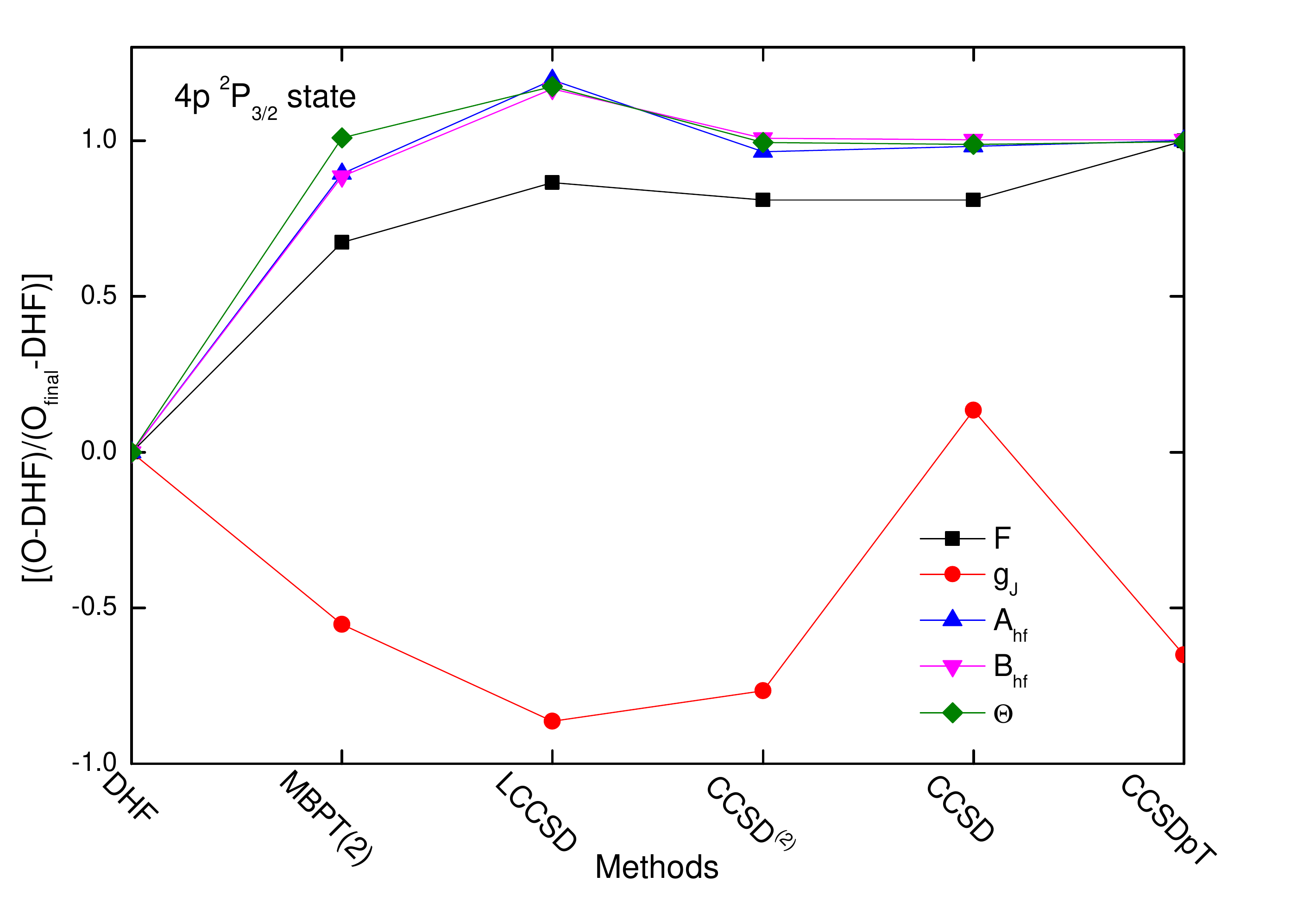}
\end{center}
\caption{(Color online) Comparison of correlation trends in the $4p ~ ^2P_{3/2}$ state of $^{43}$Ca$^+$ among the considered properties from the
employed relativistic many-body methods.}
\label{fig8}
\end{figure}

In Fig. \ref{fig3}, we plot the ($g_J-$DHF)/($g_J({\text{final}})$-DHF) values of the $g_J$ factors of the considered states from all the employed
methods. We have also subtracted the $\Delta g^{Q}_J$ values to highlight only the correlation contributions. As can be seen the correlation effects
included through the MBPT(3) method give unusually large contributions, especially in the $3d ~ ^2D_{3/2}$ and $3d ~ ^2D_{5/2}$ states. It to
be noted here is that the differences between the results from the MBPT(2) and MBPT(3) methods represent the contributions from the triples
at the lowest-order of the $T_2^{\dagger}OS_{3v}^{pert}$ RCC term. Again, the all-order triples effects are found to be crucial in Table \ref{ahfgj}
to compare the final results with the experimental values. However, in the above plot the differences in the results from the CCSD$^{(2)}$,
CCSD and CCSDpT methods are not clearly visible because of the large values plotted from the MBPT(3) method. Nevertheless, we find a completely
different trend of the correlation effects in the evaluation of the $g_J$ factors than the $A_{hf}$ values, as well as from the field shift
constants.

In Table \ref{bhfth}, we present results for $B_{hf}$ and $\Theta$ and compare them with their available experimental values. These quantities are
finite only for the states with $J > 1/2$. The most precise experimental value for $B_{hf}$ is available in the $3d ~ ^2D_{5/2}$ state. The
correlation trends for the evaluation of $B_{hf}$ are found to be exactly similar in the determination of the $A_{hf}$ values for their
corresponding states except in the $3d ~ ^2D_{5/2}$ state. The correlation trend in this state is also similar with the $4p ~ ^2P_{3/2}$ and
$3d ~ ^2D_{3/2}$ states. We have compared our calculated results with the available most precise measurements \cite{silverans,ruiz,norterhauser,benhelm}. We have quoted here two experimental values for the $4p ~ ^2P_{3/2}$ state. The reason for this is,
we had obtained $Q_{nuc}=0.0444(6)b$ \cite{sahoo1} by combining our calculation with the high precision experimental $B_{hf}$ value of the
$3d ~ ^2D_{5/2}$ state \cite{benhelm}. Our present calculation also conform with our previously reported value even after accounting more
non-linear terms in Eq. (\ref{prpeq}). However, we are unable to produce the $B_{hf}$ values close to the experimental results, especially
for the $3d ~ ^2D_{5/2}$ state, when $Q_{nuc}= 0.028(9)b$ from Ref. \cite{ruiz} is multiplied with our calculations. Considering $Q_{nuc}=0.0444(6)b$,
we are able to produce $B_{hf}$ of the $4p ~ ^2P_{3/2}$ state as $-6.693$ MHz, which is very close to the previously reported experimental value
as $-6.7(1.4)$ MHz \cite{silverans} and differ significantly from the latest measured value $-4.2(1.3)$ MHz \cite{ruiz}. Nevertheless, we get
results agreeing within the error bars of the other experimental results in all the three states. We had also discussed about other calculations
on these quantities in our earlier work \cite{sahoo1}.

\begin{figure}[t]
\begin{center}
\includegraphics[width=8.5cm,height=6.0cm]{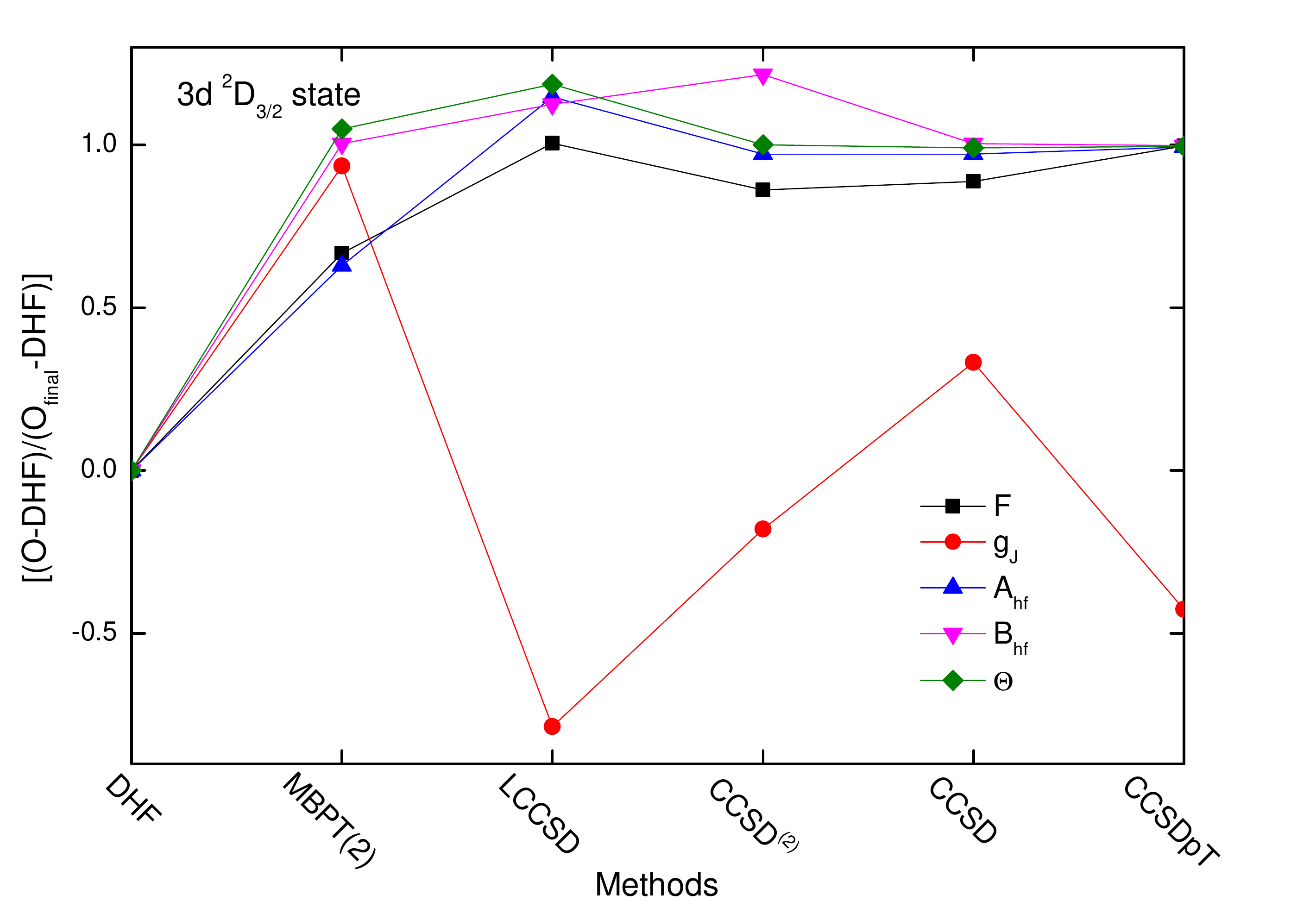}
\end{center}
\caption{(Color online) Comparison of correlation trends through the employed many-body methods for various properties in the metastable
$3d ~ ^2D_{3/2}$ state of $^{43}$Ca$^+$.}
\label{fig9}
\end{figure}

To recognize the trends of the correlation effects in the determination of the $B_{hf}$ values, we plot the ($B_{hf}-$DHF)/($B_{hf}({\text{final}})$-DHF)
values from different methods in Fig. \ref{fig4}. It shows the correlation trends in the $4p ~ ^2P_{3/2}$ and $3d ~ ^2D_{3/2}$ states are almost
similar, while they are quite large in the $3d ~ ^2D_{5/2}$ state. Differences in the results from the CCSD and CCSDpT methods indicate that
partial triples contributions in the former two states are almost negligible and it is significant in the later state. These plots are also
different than the trends observed in the evaluation of the $A_{hf}$ values in the respective states.

There is only one experimental value available for $\Theta$ of the $3d ~ ^2D_{5/2}$ state. As stated earlier, the electronic components of
$\mathcal{Q}_{el}$ and $B_{hf}$ determining operators are same and differ in their radial behavior. As seen in Table \ref{bhfth}, the correlation
trends from the DHF value seem to be completely different in the evaluation of the $B_{hf}$ and $\Theta$ values. In the case of the evaluation of
$\Theta$, the DHF method gives large values while all the considered many-body methods reduce these values due to inclusion of the correlation
effects. The all-order methods at the LCCSD method approximation gives very small values, but the CCSD method increases their values. It implies
that the electron correlation effects incorporated through the non-linear terms are found to be quite significant. Higher relativistic corrections
to these quantities are also found to be negligible. The contributions from the leading order triples are also very small in all these states. In our
previous calculation \cite{ca+bks2}, we had reported these values for the $3d ~ ^2D_{3/2}$ and $3d ~ ^2D_{5/2}$ states using the CCSD method.
Those CCSD results were much larger than the present CCSD values owing to two reasons: First of all a finite number of effective one-body terms
from Eq. (\ref{prpeq}) were considered and the self-consistent procedure improve the results notably. Moreover, it was demonstrated in
another calculation \cite{jiang} that correlation contributions from the orbitals belonging to the higher orbital angular momenta are crucial
in achieving precise values of $\Theta$ in Ca$^+$. We also found inclusion of orbitals from the orbital angular momentum beyond $l=4$, up to
which it was considered in our previous work \cite{ca+bks2}, reduce the values quite remarkably.

\begin{figure}[t]
\begin{center}
\includegraphics[width=8.5cm,height=6.0cm]{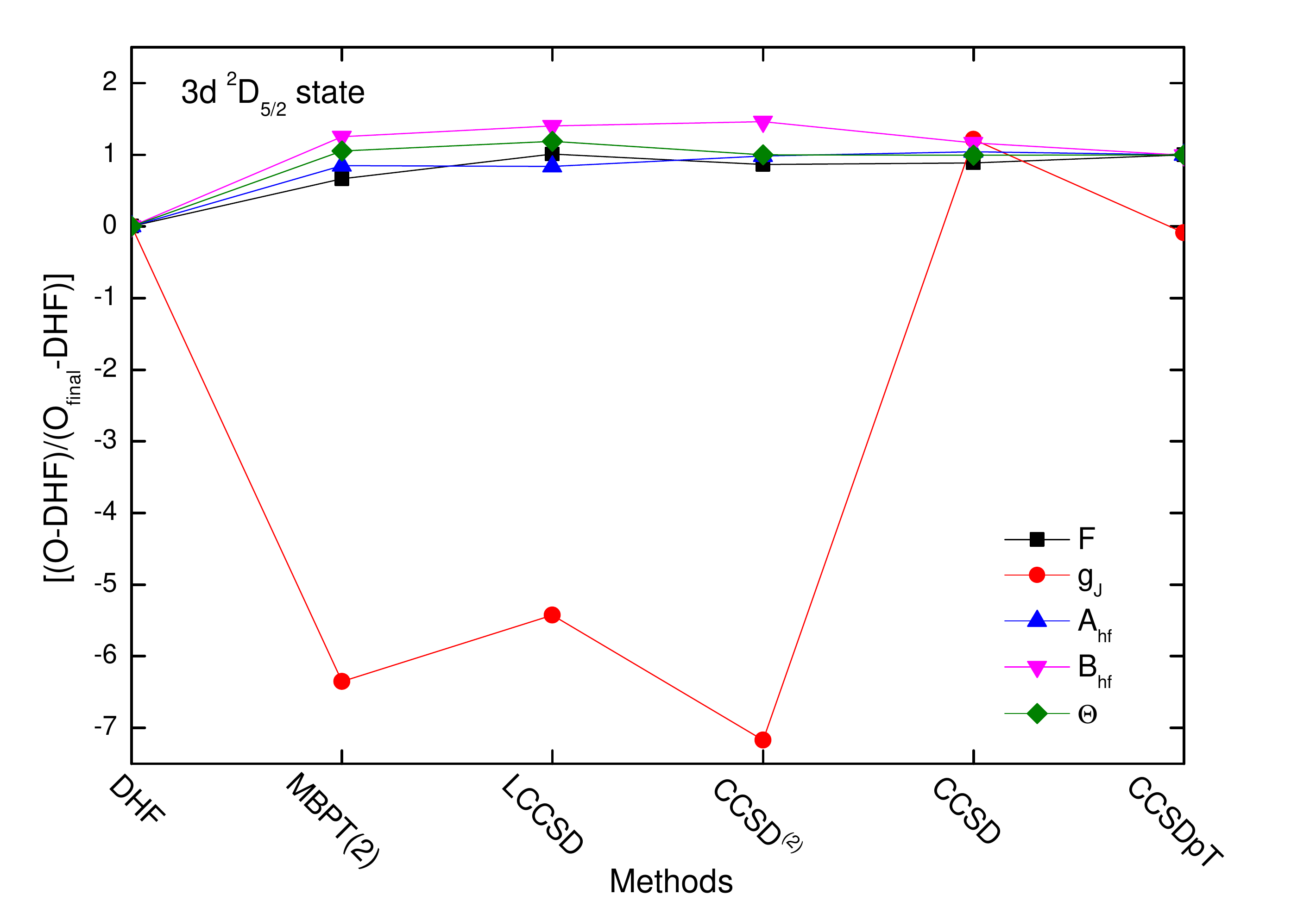}
\end{center}
\caption{(Color online) Comparison of correlation trends through the employed many-body methods for various properties in the metastable
$3d ~ ^2D_{5/2}$ state of $^{43}$Ca$^+$.}
\label{fig10}
\end{figure}

The correlation trends in the evaluation of $\Theta$ are shown in Fig. \ref{fig5} by plotting the ($\Theta-$DHF)/($\Theta({\text{final}})$-DHF) values
from all the methods in the $4p ~ ^2P_{3/2}$, $3d ~ ^2D_{3/2}$ and $3d ~ ^2D_{5/2}$ states. As can be seen from the figure, the correlation effects
are behaving almost similar trends in all the states and different than the trends observed in the evaluation of the $B_{hf}$ values. The LCCSD
method does not seem to be good approximation to evaluate this property.

 After adducing different trends of electron correlation effects in the evaluation of a given property among the considered states of Ca$^+$, we intend
now to compare these trends among different properties for a given state. For this purpose, we plot the ($O-$DHF)/($O_{\text{final}}$-DHF) values, for
the respective operator $O$, comparing all the considered properties in the $4s ~ ^2S_{1/2}$, $4p ~ ^2P_{1/2}$, $4p ~ ^2P_{3/2}$, $3d ~ ^2D_{3/2}$ and
$3d ~ ^2D_{5/2}$ states of $^{43}$Ca$^+$ in Figs. \ref{fig6}, \ref{fig7}, \ref{fig8}, \ref{fig9} and \ref{fig10}, respectively. We have not used results
from the MBPT(3) method as the correlation contributions to the $g_J$ factors from this method are very large. Thus, when results from the MBPT(3)
method are included, the correlation effects from other methods are not clearly distinguished. It is evident from these plots that
correlation trends are not similar in all the properties and they differ among the properties described by operators with the same ranks but
different radial behaviors. In Figs. \ref{fig8} and \ref{fig9}, it is also obvious that their behaviors are not unique in atomic states
having same angular momentum. We can argue by comparing Figs. \ref{fig9} and \ref{fig10} that correlation effects follow distinct trends in
the states having same parity and orbital angular momentum $l$ but different total angular momentum $J$.

\section{Summary}

We have investigated dependencies of electron correlation effects with the rank and radial behaviors of the electronic components of many
physical operators describing different atomic properties. For this purpose, we have evaluated field shift constants, hyperfine structure
constants, $g_J$ factors and electric quadrupole moments in the first five low-lying states belonging to different orbital angular momenta of
Ca$^+$. We have considered the DHF method to determine the zeroth-order results, then employed second-order
and third-order many-body perturbation theories, and linearized coupled-cluster and non-linearized coupled-cluster methods with all the singles and
doubles excitations and important triples excitations in the perturbative approaches. To demonstrate propagation of electron correlation effects
in this ion in the evaluation of various properties of the considered states, we have implemented the above many-body methods in the Fock-space
framework adopting Bloch's equations. Correlation trends in all the investigated states were compared from all the employed methods for the
individual property and then they were compared for all the properties in a given state. The following conclusions are drawn from these
analyses:

(i) We find almost all the contributions to the reported results are coming from the electron correlation effects from the Dirac-Coulomb
interactions, while the higher relativistic effects are non-negligible in the ground state.

(ii) From the estimate of leading order triples contributions, it implies that it is imperative to consider full triple excitations to achieve
very high precision results for the $g_J$ factors in the considered ion.

(iii) Even when the properties under considerations have same angular factors but differ in radial behaviors, the electron correlation effects
in a given state can exhibit very different trends in these properties.

(iv) Sometime a lower-order theory, such as the second-order many-body perturbation method, can predict more accurate results than an all-order
perturbative method, like the LCCSD method.

(v) Our analysis suggests that by just comparing theoretical values from an approximated many-body method for a particular property with the
corresponding experimental results cannot justify capability of the method to produce high-precision results for any general studies. As a solution 
to this problem, we would like to suggest that uncertainties to theoretical results can be estimated by analyzing contributions from the higher-order 
correlation effects that can arise through the neglected higher-level configurations in the calculations. In case, this leads to complication in a 
sophisticated many-body method like a truncated RCC method then accuracies of the wave functions should be adjudged by carrying out calculations 
of many properties of the atomic states having different radial and angular behaviors and comparing them with the available experimental values. 

 The above findings can be very useful while bench-marking the approximated many-body methods to reproduce the experimental results and for
providing high-precision theoretical results. It will offer some confidence on the theoretical studies of the respective properties to gauge uncertainties 
that may appear in the truncation of an employed many-body method. This will also help in judging the uncertainties of matrix elements of any given 
physical operators more reliably which cannot be obtained directly from experiments, by analyzing the radial and angular moment factors of the corresponding
operators. Knowledge of behavior of electron correlation effects in an atomic system can guide an experimentalist to use the available theoretical 
values from different many-body methods for performing measurements in the right directions prior to the observations and it can give confidence in using 
theoretical values from a method for which experiments cannot be conducted.

\section*{Acknowledgement}

C.-B. L. acknowledges support from National Science Foundation of China (Grant No. 91536102 and 91336211) and the Strategic Priority Research Program 
of CAS (Grant No. XDB21030300), and B. K. S. acknowledges financial support from CAS through the PIFI fellowship under the project number 2017VMB0023. 
Computations were carried out using Vikram-100 HPC cluster of Physical Research Laboratory (PRL), Ahmedabad, India.

\end{document}